\documentclass[12pt]{article}
\usepackage{caption}
\usepackage{subcaption}
\usepackage{booktabs}
\usepackage{graphicx}
\usepackage{tikz}
\usepackage{algorithm, algpseudocodex}
\usepackage{multirow, multicol}
\usepackage{changepage}
\usepackage{enumitem}
\usepackage{natbib}
\usepackage{url} 

\usepackage{amsmath, amsfonts, amssymb, amsthm}

\newcommand{\bi}{\begin{itemize}}
\newcommand{\ei}{\end{itemize}}

\newcommand{\be}{\begin{enumerate}}
\newcommand{\ee}{\end{enumerate}}

\newcommand{\norm}{\mathrm{N}}

\newcommand{\Dcal}{\mathcal{D}}
\newcommand{\Gcal}{\mathcal{G}}

\newcommand{\Zcal}{\mathcal{Z}}

\newcommand{\1}{\mathbf{1}}

\newcommand{\D}{\mathbf{D}}

\newcommand{\E}{\mathbf{E}}

\newcommand{\G}{\mathbf{G}}

\newcommand{\I}{\mathbf{I}}

\newcommand{\N}{\mathbf{N}}

\newcommand{\vv}{\mathbf{v}}
\newcommand{\V}{\mathbf{V}}
\newcommand{\U}{\mathbf{U}}

\newcommand{\X}{\mathbf{X}}

\newcommand{\y}{\mathbf{y}}

\newcommand{\z}{\mathbf{z}}

\newcommand{\mubf}{\boldsymbol{\mu}}

\newcommand{\xibf}{\boldsymbol{\xi}}

\newcommand{\thetabf}{\boldsymbol{\theta}}

\newcommand{\etabf}{\boldsymbol{\eta}}

\newcommand{\sigmabf}{\boldsymbol{\sigma}}
\newtheorem{definition}{Definition}[section]
\newtheorem{proposition}{Proposition}[section]
\newtheorem{theorem}{Theorem}[section]
\newtheorem{lemma}{Lemma}[section]
\newtheorem{corollary}{Corollary}[section]

\newcommand{\blind}{0}

\begin{document}

\def\spacingset#1{\renewcommand{\baselinestretch}%
{#1}\small\normalsize} \spacingset{1}


\if0\blind
{
  \title{\bf Mixture of Directed Graphical Models for Discrete Spatial Random Fields}
  \author{J. Brandon Carter\thanks{
    The authors gratefully acknowledge support from the Eunice Kennedy Shriver National Institute on Child Health and Human Development (R01HD088545 and P2CHD042849) and the National Science Foundation (2502935)}\\
    The Hershey Company \\[.25cm]
    and \\[.25cm]
    Catherine A. Calder\footnote{catherine.calder@austin.utexas.edu} \\
    Department of Statistics and Data Sciences \\ University of Texas at Austin}
  \maketitle
} \fi

\if1\blind
{
  \bigskip
  \bigskip
  \bigskip
  \begin{center}
    {\LARGE\bf Mixture of Directed Graphical Models for Discrete Spatial Random Fields}
\end{center}
  \medskip
} \fi


\begin{abstract}
Current approaches for modeling discrete-valued outcomes associated with spatially-dependent areal units incur computational and theoretical challenges, especially in the Bayesian setting when full posterior inference is desired. 
As an alternative, we propose a novel statistical modeling framework for this data setting, namely a mixture of directed graphical models (MDGMs).
The components of the mixture, directed graphical models, can be represented by directed acyclic graphs (DAGs) and are computationally quick to evaluate.
The DAGs representing the mixture components are selected to correspond to an undirected graphical representation of an assumed spatial contiguity/dependence structure of the areal units, which underlies the specification of traditional modeling approaches for discrete spatial processes such as Markov random fields (MRFs).
Notably, the MDGM is not proposed as an approximation to an MRF, but rather as an alternative that provides valid posterior inference while being computationally faster than exact MRF inference and more principled than the pseudo-likelihood approximation (aMRF) commonly used in practice.
We introduce the concept of compatibility to show how an undirected graph can be used as a template for the dependencies between areal units to create sets of DAGs which, as a collection, preserve the dependencies represented in the template undirected graph.
Lastly, we compare highlighted classes of MDGMs to MRFs and a popular Bayesian MRF model approximation used in high-dimensional settings in a series of simulations and an analysis of ecometrics data collected as part of the Adolescent Health and Development in Context Study.  
\end{abstract}

\enlargethispage{2\baselineskip}
\noindent%
\vfill

\newpage
\spacingset{1.75} 
\section{Introduction}
\label{sec:intro}
Statistical models for spatially-dependent, discrete-valued observations associated with areal units (i.e., discrete space) are used routinely in areas of application ranging from disease mapping to small area estimation to image analysis.
As ``discrete'' is used to describe multiple features of the analytic setting, we standardize our use of the term.
We use \textit{discrete-valued} observations/latent variables to refer to the observations/latent variables associated with areal units that take on discrete values.
We use \textit{discrete spatial process} to refer generally to the areal unit/lattice setting.  Despite a rich literature on both classical and Bayesian models for this setting, theoretical and computational considerations remain, particularly in the Bayesian setting when complete posterior inference is desired. 
To address the computational and theoretical concerns of existing models (described below), we propose a novel framework for modeling spatial dependence of observations associated with a collection of $n$ areal units which together form a nonoverlapping partition of the spatial/geographic domain. 
The areal units may be a regular grid/lattice or be an irregular partition (e.g., administrative districts). 
Without loss of generality and to motivate our new framework, we assume at each areal unit there is a collection of zero-one binary observations, $[y_{i1},\dots,y_{im_{i}}]$, where $m_i$ is the number of observations at areal unit $i$. 
Additionally, we assume that there is a single binary latent variable, $z_i$, associated with each areal unit for $i=1,\dots,n$.
Let $\y=[[y_{11},\dots,y_{1m_{1}}], \dots,[y_{n1},\dots,y_{nm_{n}}]]$ denote the complete collection of observations and $\z=[z_1,\dots,z_n]$ the vector of binary latent variables. 

Standard spatial statistical models for areal data assume that there is a known \textit{natural undirected graph} (NUG, our terminology) with vertices/edges denoting the areal units and the spatial proximity/contiguity of pairs of areal units, respectively.
\textit{Markov random fields} (MRFs) are a default choice for prior distribution on $\z$, as an MRF, defined for a given NUG, is in fact a probability distribution whose conditional distributions elicit the same dependence structure (i.e. dependencies between areal units) as the specified NUG \citep{besag1974, besag1975}. 
While the one-to-one correspondence between a NUG and an MRF leads to a simple and intuitive specification of the joint probability distribution of $\z$, inference on the parameters governing an MRF is computationally burdensome due to an intractable normalizing constant when $\z$ is discrete. 
While diverse approximation methods for the MRF have been proposed \citep{reeves_pettitt2004, friel_etal2009, mcgrory_etal2009, mcgrory_etal2012}, the approach that uses the \textit{pseudo-likelihood} of \citet{besag1975} in place of the analytical form of the intractable MRF prior density persists in Bayesian analyses \citep{heikkinen_hogmander1994, hoeting_etal2000, wang_etal2000, hughes_etal2011, pereyra_mclaughlin2017, marsman_haslbeck2023}.
While relatively easy to implement and computationally efficient, this approach does not guarantee valid posterior inference (see Section~\ref{sec:pl_approx}).

As an alternative to specifying a joint probability model for the observed and latent variables associated with the areal units based on the assumed NUG directly, we propose using the NUG as a template for the dependencies that exist between the areal units.
From the NUG, we derive a collection of \textit{directed acyclic graphs} (DAGs) which is used to specify probability distributions, \textit{directed graphical models} (DGMs), that together capture the spatial dependence between random variables associated with the areal units.
Directed graphical models are also called Bayesian networks in older literature, a name we dislike as there is nothing inherently Bayesian about a directed graphical model.
The main advantage of a DGM is the straightforward factorization of the probability distribution into conditional distributions, which leads to quick computational evaluation of a DGM (Section~\ref{sec:priors}).
Let $\Dcal(\N) = \{\D_1, \dots, \D_L\}$ be a set of DAGs which honor the dependencies of the NUG, $\N$, where $L$ equals the cardinality of the set and $\N$ and $\D$ define the structure of a NUG and DAG, respectively, through a set of vertices and edges; we give formal definitions of $\N$ and $\D$ in Section~\ref{sec:nug_as_template}.
In order to formally describe how a DAG can be derived from a template NUG which honors the topology of the NUG, we define the notion of \textit{weak compatibility} in Section~\ref{sec:weak_compatibility}.
In Section~\ref{sec:strong_compatibility}, we provide a more robust definition of compatibility, which also ensures fidelity of the DAG to the conditional independence relationships of the NUG.
Specific classes of DAGs that can be used in the mixture are given in Section~\ref{sec:st} (spanning trees) and Appendix~\ref{ap:classes_of_dags} (acyclic orientation and rooted graphs).
Given the set of DAGs, $\Dcal$, and a DGM for $\z$, written as $p(\z|\D,\xibf)$, that depends on the structure of the DAG and parameters $\xibf$, we specify the prior on $\z$ as
\begin{equation}
\label{eq:mixture_dgm}
  p(\z|\xibf) = \sum_{\D \in \Dcal(\N)} p(\D)p(\z|\D,\xibf),  
\end{equation}
a \textit{mixture of directed graphical models} (MDGM), where the mixture weights are given implicity by the prior over the collection of DAGs, $p(\D)$ for $\D\in\Dcal$. 
Given the latent $\z$, we model each $y_{ij}$ as conditionally independent Bernoulli random variables, 
$$
p(\y|\z,\etabf) = \prod_{i=1}^n \prod_{j=1}^{m_i}\eta_{z_i}^{y_{ij}}(1-\eta_{z_i})^{1-y_{ij}},
$$
with noise parameters $\etabf=[\eta_0, \eta_1]$, where $\eta_0$ is the probability that $y_i=1$ when $z_i=0$ and $\eta_1$ is the probability that $y_i=1$ when $z_i=1$.
In this setting the $y_i$'s can be viewed as a noisy version of the true underlying binary image $\z$.
A realization $\z$ of the spatial field with respect to the NUG is called a \textit{configuration} of the graph. 
We call realizations from both MRFs and MDGMs configurations of the NUG. 
While continuous-valued distributions for $\y$ given the latent $\z$ are more common in the literature, we focus on this discrete outcome setting with spatial dependence introduced through a discrete latent variable as fully Bayesian approaches in this setting have been less explored \citep{arnesen_tjelmeland2015}.
We note, however, the ideas presented in the MDGM framework are generally applicable to the variety of combinations of discrete or continuous latent and observed variables.
Additionally, while we present the MDGM as a prior in a hierarchical framework, the MDGM can also serve as a direct data model for discrete-valued outcomes associated with areal units.

The main contributions of the paper are as follows.
In Section \ref{sec:nug_as_template}, we define the notion of compatibility, with its weak and strong versions, which provides a principled way to construct a DAG (and the corresponding DGM) from a template NUG.
Through compatibility, sets of DAGs can be specified which are distinguished by some feature of the DAGs in the set, e.g. minimally connected DAGs, namely spanning trees, or fully connected DAGs, namely acyclic orientations of the NUG.
We highlight the class of spanning trees, which are strongly compatible with any NUG and discuss model fitting via MCMC in Section~\ref{sec:mod_fit}. Additional MDGM classes are presented in the Appendix.
Compatibility expands the statistical toolkit by linking relevant graph theory to statistical modeling and by providing a formal foundation for the development of new classes of DAGs to be used in the MDGM, as specified in Section~\ref{sec:priors}. 
Our notion of compatibility unifies existing approaches which have sought to utilize DAGs and other graph structures to facilitate inference, though often in a different inferential setting such as structure learning \citep{wu_doerschuk1995, meila_etal1997, meila_jordan2001, pletscher_etal2009, thiesson_etal1999}.
We compare our models, via simulation and data analysis (Sections~\ref{sec:mod_eval} and \ref{sec:phys_dis_example}), using an MDGM prior to corresponding models with an MRF prior.
In addition, we include a comparison to the widely-used, but theoretically ungrounded, ``approximate MRF prior" (aMRF), discussed in Section~\ref{sec:pl_approx}.

\section{The NUG as a Template for a DAG}
\label{sec:nug_as_template}

In this section, we introduce the concept of compatibility, which is the criteria used to determine whether a DAG or set of DAGs honor the topology of the NUG.
DAGs which honor the structure of the NUG are said to be weakly compatible, while DAGs which honor the dependence relationships implied by the topology of the NUG are strongly compatible. 
We formally define weak and strong compatibility below.

NUGs and DAGs are each graphical representations of probability models, MRFs and DGMs, respectively.
We address specification of an MRF given a NUG and a DGM given a DAG
in Section~\ref{sec:priors}; for now, we begin with weak compatibility and focus on how the topology of the NUG serves as a template for the structural dependencies between vertices.
To this end we introduce some notation. 

Without loss of generality, we center discussion on the NUG as the template for sets of DAGs, though the concepts apply to any undirected graph. 
Let $\N=\{\V,\bar\E\}$, be a NUG with set of vertices, $\V=\{i:i=1,\dots,n\}$, where $n$ is the number of areal units, and set of unordered pairs, $\bar\E$, where the unordered pair, $\{i,j\}$, denotes that there is an undirected edge between vertices $i$ and $j$.
In a NUG, we say that vertices $i$ and $j$ are neighbors if $\{i,j\}$ is an element of $\bar\E$. 
We introduce the notation
$$
\partial(i) = \{j:j\text{ is a neighbor of } i\},
$$
to represent the set of neighbors of vertex $i$ and note that $\partial(i) = \{j:\{i,j\}\in\bar\E\}$.
The NUG in the discrete spatial setting generally takes one of two forms, a \textit{first} or \textit{second-order} neighborhood structure (see Appendix~\ref{ap:nugs}), where neighbor relationships are defined by the contiguity of areal units. 

A key feature of the NUG is that the undirected edge, $\{i,j\}$, represents a symmetric relationship between vertices $i$ and $j$.
In contrast, a DAG, defined on the same set of vertices and denoted $\D=\{\V,\vec\E\}$, has an edge set, $\vec\E$, comprised of ordered pairs of vertices $(i,j)$, which we use to indicate, somewhat unconventionally, that there is a directed edge from $j$ to $i$.
For the directed edge $(i,j)$ we say that $j$ is a parent of $i$ and define
$$
\pi(i) = \{k:k\text{ is a parent of }i\}
$$
to be the set of all parents of vertex $i$.
Alternatively, for the ordered pair $(i,j)$, $i$ is a child of $j$.
DAGs are the subset of directed graphs which do not contain any directed cycles.
The definition of a cycle relies on the definition of a path.
A path is any sequence of connected edges (without regard to direction) and a directed path is a sequence of directed edges where the child of the preceding edge is the parent of the next edge in the sequence. 
A directed cycle is a sequence of ordered pairs, $(i,j), (k,i),\dots,(j,l)$, which start at vertex $j$ and follow a directed path back to $j$.
Lastly, let $\G=\{\V,\E\}$, denote a general graph, where the edge set $\E$ can be comprised of both ordered and unordered pairs of vertices.
A graph $\G'=\{\V',\E'\}$ is called a subgraph of $\G$, denoted $\G'\subseteq\G$, if $\V'\subseteq\V$ and $\E'\subseteq\E$.

\subsection{Weak compatibility of a graph to a NUG}
\label{sec:weak_compatibility}
How the NUG serves as a template of structural dependencies becomes clear with an alternative representation of the graph structure.
Let $A(\G)$ be a matrix representation of the graph $\G$.
This \textit{adjacency matrix}, $A(\G)$, is constructed by setting the $(i,j)$ and $(j,i)$ positions of the matrix to one if the undirected edge, $\{i,j\}$, is in the edge set and zero otherwise.
For directed edges, $(i,j)$, in the edge set, only the $(i,j)$ position is set to one.
By definition, the matrix $A(\N)$ is symmetric. 
A property of the adjacency matrix for a DAG, $\D$, is that there exists a permutation of the rows and columns of $A(\D)$ for which $A(\D)$ is lower triangular. 
Additionally, in a DAG, the $i$th row of $A(\D)$ is a $m$-dimensional vector with binary indicators for the parents of $i$, that is $\pi(i)$.
Now we can define weak compatibility of a graph with the NUG.
\begin{definition}
\label{def:weak_compatibility}
A graph, $\G=\{\V,\E\}$, with vertices, $\V$, is weakly compatible with a NUG, $\N=\{\V,\bar\E\}$, defined on the same set of vertices, denoted $\G\bumpeq \N$, when $(i,j)$ is equal to one in $A(\G)$ implies that $(i,j)$ is equal to one in $A(\N)$.
\end{definition}
See Appendix~\ref{ap:compatible_example} for examples of weakly compatible and incompatible DAGs for a given NUG.
Note, that compatibility does not require that $|\E|=|\bar\E|$, where $|\X|$ denotes the cardinality of a set $\X$. We can extend this definition to sets of graphs. 
Let $\Gcal = \{\G_1, \dots, \G_k\}$ be a set of graphs.
Then the set, $\Gcal$, is weakly compatible with $\N$, denoted $\Gcal\bumpeq\N$, if for each $\G\in\Gcal$, $\G$ is weakly compatible with $\N$. 


In the following section, we introduce strong compatibility and show that strongly compatible graphs are a subset of weakly compatible graphs. 
An alternative definition of weak compatibility is useful to this end and relies on the notion of a skeleton graph.
The skeleton of a graph $\G$, denoted $\U(\G)=\{\V,\{\{i,j\}: (i,j)\in\E\text{ or }\{i,j\}\in\E\}\}$, is the undirected graph for which all directed edges of $\G$ become undirected.
The following lemma provides an equivalent definition of weak compatibility.
\begin{lemma}
\label{lemma:subgraph}
    A graph $\G$ is weakly compatible with $\N$ if and only if the skeleton of $\G$, $\U(\G)$, is a subgraph of $\N$.
\end{lemma}
From Definition~\ref{def:weak_compatibility} and Lemma~\ref{lemma:subgraph}, graphs weakly compatible with a NUG cannot have edges between vertices that are not neighbors in the NUG. 
Strong compatibility tightens the structural criterion of weak compatibility further by considering conditional dependence relationships induced by directed edges. 

\subsection{Strong compatibility of a graph to a NUG}
\label{sec:strong_compatibility}
The appeal of graphical models is their representation of the dependence relationships between sets of variables in a probability distribution through a graph. 
In Appendix~\ref{ap:ci} we review how DAGs and undirected graphs can each express different sets of conditional independence relationships of the form $\z_A\perp \z_B|\z_C$ for sets of variables indexed by the sets vertices, $A,B,C\in\V$, in the vertex set. 
Let $I(\G)$ denote the set of conditional independence relationships of the form $A\perp B|C$ asserted by the graph $\G$.
To build up to our definition of strong compatibility which relates the sets $I(\N)$ and $I(\D)$ to each other, we first provide some common terms describing relationships between the sets of conditional independence relationships asserted by a graph and a probability distribution.
Let $I(p)$ be the set of conditional independence relationships of the form, $A\perp B|C$, that hold in the probability distribution $p$; that is, $\z_A\perp \z_B|\z_C$.
We say that $\G$ is an \textit{independence map} (I-map) of $p$ if $I(\G) \subseteq I(p)$; expressly, $\G$ does not assert any conditional independence relationships that are not present in the probability distribution \citep{pearl1988probabilistic}.
When $I(p) \subseteq I(\G)$, then we say that $\G$ is a \textit{dependence map} (D-map) of $p$, accordingly, $\G$ does not assert any conditional \textit{dependencies} that are not present in the probability distribution. 
When $I(\G)=I(p)$, then $\G$ is a \textit{perfect map} (P-map) of $p$.

We extend these definitions to relate two graphs to each other. For $\G$ and $\G'$ defined on the same set of vertices, we say that $\G$ is a I-map of $\G'$ or, equivalently, $\G'$ is an D-map of $\G$ when $I(\G)\subseteq I(\G')$. 
When $I(\G)=I(\G')$, then $\G$ and $\G'$ are said to be Markov equivalent \citep{andersson1997characterization}.
The definition of strong compatibility is as follows. 

\begin{definition}
\label{def:strong_compatibility}
A graph, $\G=\{\V,\E\}$, with vertices, $\V$, is strongly compatible with a NUG, $\N=\{\V,\bar\E\}$, defined on the same set of vertices, denoted $\G\Bumpeq \N$, when $I(\N) \subseteq I(\G)$. That is, $\G$ is a D-map of $\N$.
\end{definition}

As with weak compatibility, we say that a set of graphs, $\Gcal$, is strongly compatible with $\N$, denoted $\Gcal \Bumpeq \N$, when for each $\G\in\Gcal$, we have $\G\Bumpeq\N$. 
The strong compatibility criterion ensures that compatible graphs do not introduce any dependencies between vertices that are not already present in the NUG. 
As a consequence, we have the following theorem with the proof in Appendix~\ref{ap:proof:strongweak}.

\begin{theorem}
\label{thm:strongweak}
    Any graph $\G$ that is strongly compatible with a NUG $\N$, $\G\Bumpeq\N$, is also weakly compatible with $\N$, $\G\bumpeq\N$. 
\end{theorem}



A DAG is weakly compatible as long as all of its directed edges correspond to undirected edges in the NUG.
The complete set of DAGs weakly compatible with the NUG, $\Dcal(\N)=\{\D:\D\bumpeq\N\}$, comprises of the combined set of all DAGs that can be derived from each possible undirected subgraph of the NUG.
Strong compatibility narrows the set of weakly compatible DAGs that simply honor the structure of the NUG to the set of DAGs which also honor the set of conditional independencies implied by the NUG.
Conditional dependencies not present in the NUG in weakly compatible DAGs are introduced by v-structures. 
A v-structure is a pattern in a graph where two directed edges meet head-to-head, as in $i\rightarrow j \leftarrow k$.
The parents of $j$, vertices $i$ and $k$, are called \textit{unprotected} or \textit{unwed} if there is no edge, directed or undirected, between $i$ and $k$.
For such a pattern we have $i\perp k$, but $i\not\perp k|j$. 
Another characterization of strong compatibility is a weakly compatible DAG that also does not contain any v-structures with ``unwed'' parents.
Trivially, the graph without any edges is strongly compatible with any NUG over the same set of vertices, though in the spatial setting such a graph is useless in learning and accounting for spatial dependence.

Since the sets of all weakly or all strongly compatible graphs are both quite large, we define a \textit{class} of compatible DAGs as a subset of the complete set of compatible DAGs that are further distinguished by specific criteria. 
One criterion is DAGs that preserve all of the edges in the NUG. Such DAGs are called \textit{acyclic orientations} with respect to the NUG and are a class of weakly compatible DAGs. 
We expound acyclic orientations and an additional class of DAGs, rooted graphs, in Appendix~\ref{ap:classes_of_dags}.
In the next section, we explore the class of spanning trees.

\subsection{Spanning Trees}
\label{sec:st}
\textit{Spanning trees} are fully connected graphs (i.e. graphs with a path between any two vertices) with a minimal edge set.
In order to be fully connected, spanning trees have $n-1$ edges.
Spanning trees are a unique class of compatible graphs: both undirected spanning trees (or simply, spanning trees) and a special case of \textit{directed spanning trees} represent equivalent sets of conditional independence relationships.
A directed spanning tree is a DAG with a spanning tree skeleton.
\textit{Rooted spanning trees} are a special case of directed spanning trees that have a single orphan vertex which is called the root of the graph (a generic DAG is also called rooted if it has a single orphan vertex).
A rooted spanning tree does not contain any v-structures because all edges are oriented away from the root giving each non-root vertex only a single parent.
Let $\Dcal^{\text{ST}}(\N)$ be the set of all possible rooted spanning trees.

Any two graphs that have the same unwed v-structures and skeleton graph are Markov equivalent \citep{verma1990equivalence}.
Accordingly, a spanning tree is Markov equivalent with any rooted spanning tree that has the same skeleton.
The Markov equivalence of a spanning tree and any rooted spanning tree with an identical skeleton has implications for statistical learning which we explain in Section~\ref{sec:priors}.
Due to the Markov equivalence, we call the class simply spanning trees, but leverage rooted spanning trees for unified computation in and presentation of the MDGM framework.

Spanning trees are also a special class of compatible graphs, which we formalize in the following theorem.

\begin{theorem}
\label{thm:spanning_trees}
    A spanning tree, $\U=\{\V,\bar\E'\}$, derived from a NUG, $\N=\{\V,\bar\E\}$, is always strongly compatible with $\N$.
\end{theorem}

The proof is provided in Appendix~\ref{ap:proof:spanning_trees}. Rooted spanning trees are also strongly compatible with the NUG because of the Markov equivalence of rooted and undirected spanning trees.
That is, for all $\D\in\Dcal^{\text{ST}}(\N)$ we have $I(\N) \subseteq I(\D)=I(\U(\D))$.
Rooted spanning trees honor the dependence relationships of the NUG because they do not contain unprotected v-structures.
This naturally leads to the following corollary of Theorem \ref{thm:spanning_trees}. 

\begin{corollary}
\label{cor:st_strong_first_order}
    For a NUG, $\N$, with first-order dependence structure on a regular lattice, a spanning tree is the only fully connected graph structure that is strongly compatible with $\N$. 
\end{corollary}

The proof is in Appendix~\ref{ap:proof:st_strong_first_order}.
The strong compatibility of spanning trees makes the class a good default option for specifying an MDGM model. We explain how to specify a MDGM model given a general set of DAGs in Section~\ref{sec:mdgm} and highlight the spanning tree class. 
In Appendix~\ref{ap:st_wilsons_size}, we describe how to generate a rooted spanning tree and how to count the size of class. 

\section{Hidden Discrete-Valued Spatial Dependence}
\label{sec:priors}

We now complete the specification of our hierarchical model,
\begin{equation}
\label{eq:general_posterior_valid}
p(\z,\thetabf,\etabf|\y) \propto p(\y|\z,\etabf)p(\z|\thetabf)p(\thetabf)p(\etabf),
\end{equation}
by specifying the priors $p(\etabf)$ and $p(\thetabf)$.
Later in this section, we define the form of the prior $p(\z|\thetabf)$ for three different cases, the MDGM, MRF and aMRF. 
For each type of prior on the latent variable, we note whether the posterior distribution is valid. 

A simple prior for $\etabf$ is, 
$$
p(\etabf) \propto p(\eta_0)p(\eta_1)I(\eta_1>\eta_0),
$$
where $I(\cdot)$ is the indicator function and the identifiability constraint, $\eta_1>\eta_0$, avoids the label switching problem of the latent variable.
In the case of the MDGM, $\thetabf=\{\D,\xibf\}$, whereas $\thetabf=\xibf$ for the MRF and aMRF. 
Let $p(\thetabf)=p(\D)p(\xibf)$ for the MDGM and $p(\thetabf)=p(\xibf)$ for the models with MRF and aMRF priors.
We use $\xibf$ to denote the spatial dependence parameters of the three different spatial models, and while the mathematical forms of the three latent variable priors that we use for analyses later in the paper are analogous, the mathematical interpretation of $\xibf$ is distinct between the MRF/aMRF and MDGM priors and across our classes of the MDGM highlighted in the paper and appendices.
In the MRF/aMRF prior, spatial dependence between the vertices of a NUG is  determined solely through the parameters $\xibf$, whereas in the MDGM classes, spatial dependence is determined through the DAGs included in the mixture and the parameters $\xibf$. 
A default choice for the prior $\D$, is a uniform prior, $p(\D)=\frac{1}{|\Dcal^\dagger(\N)|}$ for $\D\in\Dcal^\dagger(\N)$, where $\Dcal^\dagger(\N)\subseteq\Dcal(\N)$ is a specific class/subset of all weakly compatible DAGs.

\subsection{MDGM Prior}
\label{sec:mdgm}
As review, a MDGM prior is specified in three steps.
First, define a NUG for the areal units, often a first or second-order neighborhood structure.
Second, select a subset of compatible DAGs to include in the mixture. 
Third, we specify the form of the DGM
\begin{equation}
\label{eq:dgm_factorization}
p(\z|\D,\xibf) = \prod_{i=1}^n p(z_i|\z_{\pi(i)},\xibf),
\end{equation}
whose factorization into conditional distributions, $p(z_i|\z_{\pi(i)},\xibf)$, is determined by the topology of the DAG. 
This factorization leads to quick evaluation of $p(\z|\thetabf)=p(\z|\D,\xibf)$, a major computational advantage for MCMC-based inference over an MRF specification for the joint distribution of $\z$.


Additionally, for a general DAG
we set
\begin{equation}
\label{eq:parent_cond_dgm}
p(z_i|\z_{\pi(i)},\beta) = \frac{\exp\left(\beta\sum_{j \in \pi(i)}\I(z_i = z_j)\right)}{\exp\left(\beta\sum_{j \in \pi(i)}\I(z_j=0)\right)  + \exp\left(\beta\sum_{j \in \pi(i)}\I(z_j=1)\right)},
\end{equation}
with $\sum_{j \in \pi(i)}\I(z_j = x) \equiv 0$ for $x=0,1$ if $\pi(i)$ is equal to the null set.
This specific form of the conditional distributions is analogous to the standard MRF specification (Section~\ref{sec:mrf}) in spatial statistics.
The above prior specification yields the prior full conditionals
\begin{equation}
\label{eq:dgm_full_cond_prior}
p(z_i|\z_{-i},\beta) \propto
\frac{\exp\left(\beta\sum_{j\in\partial(i)}\I(z_i = z_j) \right)}{\prod_{k\in\kappa(i)}\left[\exp\left(\beta\sum_{j\in\pi(k)}\I(z_j = 0)\right) + \exp\left(\beta\sum_{j\in\pi(k)}\I(z_j = 1) \right)\right]},
\end{equation}
where we define $\partial(i)=\{\pi(i),\kappa(i)\}$ for a DAG and let $\z_{-i}$ be the vector of latent variables with the $i$th variable removed. 
For the ordered pair $(i,j)$, we say $i$ is a child of $j$, and let
$$
\kappa(i)=\{k:k\text{ is a child of }i\}
$$ 
be the set of all children of vertex $i$.
From the prior full conditionals we see that $p(z_i|\z_{-i})$ depends only the set of variables associated with the vertices $\{\pi(i), \kappa(i), \pi(\kappa(i))\}$, that is, the set of parents of $i$, children of $i$, and the parents of the children of $i$.


We note that the posterior, $p(\z,\D,\xibf,\etabf|\y)$, for a general MDGM as defined by Equation~\ref{eq:general_posterior_valid}, is a valid probability distribution as each $p$ on the right hand side is a valid probability distribution.
The particular form of the prior $p(\z|\D,\xibf)$ provides quick evaluation, needed for updates on $\D$ and $\xibf$ in the MCMC algorithms described in Section~\ref{sec:mod_fit}. 
We now highlight the spanning tree class.

\subsubsection{Spanning Tree Class}
\label{sec:mdgm_st_prior}
First, we specify a prior for the spanning tree class. 
As noted above when describing the loop erased random walk algorithm, the probability of a spanning tree is proportional to the product of the weights of its edges.
Specifying uniform weights for the edges of the NUG, \textit{a priori}, yields the uniform prior $p(\D) = \frac{1}{|\Dcal^{\text{ST}}(\N)|}$, over all possible spanning trees in the set $\Dcal^{\text{ST}}(\N)$.
Prior information on edge inclusion probabilities can be included into the model by assigning different weights to the edges in the NUG. Also of note, in the prior full conditionals for the spanning tree class, the denominator of Equation~\ref{eq:dgm_full_cond_prior} becomes a constant, $2(1+\exp(\beta))^{n-1}$, because a rooted spanning tree does not have any unprotected v-structures.

Consider now the characterization of spatial dependence in the spanning tree class. 
The pairwise spatial dependence between two vertices is determined by the proportion of times the two vertices are connected across all spanning trees in the mixture and the spatial dependence parameter $\beta$.
We can think of $\beta$ as the global spatial dependence parameter and the weight edges as local adjustments to spatial dependence.
As $\beta$ increases, the areal units will be more clustered. 
The probability that a particular edge appears in any given spanning tree in the mixture is proportional to the weight of that edge.
With uniform prior weights on the edges in the spanning tree class, \textit{a priori}, all edges have equal probability of appearing in the mixture, thus there are no local adjustments to spatial dependence.
For nonuniform edge weights, as in the posterior distribution, local adjustments to the degree of dependence do occur, making the spanning tree class very flexible in representing spatial dependence. 

Lastly, we address the previously alluded to property of the spanning tree class in regard to identifiability of the DAG topology. 
It can be shown that specifying a probability distribution for a tree graph can be written equivalently as an MRF or as a DGM, where the DGM is represented by a rooted spanning tree \citep{meila_jordan2001}. 
Since the DGMs corresponding to the $n$ different rooted spanning trees which can be created from an undirected tree are Markov equivalent, the root of the rooted spanning tree is unidentifiable, thus we are only able to learn the posterior distribution of the undirected tree structures. 
As a result, this class is less strictly a subset of DAGs when compared to the other two classes of DAG models that we present in the appendix; however, as a DGM and MRF specification for the same tree structure are equivalent, we can still utilize Equations~\ref{eq:dgm_factorization} and \ref{eq:parent_cond_dgm} for the spanning tree class to maintain unified computation between the MDGM classes.



\subsection{MRF Prior}
\label{sec:mrf}
We now present the MRF prior to which we will compare our new MDGM framework. 
As stated in the introduction, MRFs are popular because of the intuitive representation of the probability distribution through the NUG.
In particular, undirected graphs satisfy the local Markov property which states that a vertex, $i$, is independent of all other vertices given its neighbors. (See Appendix~\ref{ap:ci} for a more thorough discussion.)
From the local Markov property, the full conditionals, $p(z_i|\z_{-i},\xibf)$, from an MRF specification reduce to $p(z_i|\z_{\partial(i)},\xibf)$.
While an MRF is defined as a probability distribution whose conditional distributions define a NUG \citep[pg. 415]{cressie1993}, we can, in actuality, start with a collection of full conditional distributions consistent with a NUG, $p(z_i|\z_{\partial(i)},\xibf)$, and obtain the corresponding MRF, which we are guaranteed is a valid joint probability distribution by the Hammersley-Clifford Theorem, for which an extended proof given by \citet{besag1974}.

The result depends on the notion of a clique.
Any single node or set of vertices which are all mutually neighbors are defined to be a clique.
Let $C=\{c_1,\dots,c_k\}$ be the set of all cliques in the graph, $\N$. 
From the Hammersley-Clifford Theorem, if the probability distribution of an MRF, $p(\z|\xibf)$ for $\z\in\Zcal$, with respect to graph $\N$ and cliques $C$, satisfies $p(\z|\xibf)>0$ for all $\z\in\Zcal$, then we can write
\begin{equation}
\label{eq:hc_thm}
p(\z|\xibf) = \frac{1}{R(\xibf)} \prod_{c\in C} \psi_c(\z_c|\xibf_c).
\end{equation}
Here $\psi_c(\z_c|\xibf_c)$ are arbitrarily chosen, nonnegative, and finite functions parameterized by $\xibf=\{\xibf_c:c\in C\}$ and the partition function,
$$
R(\xibf) = \sum_{z\in\Zcal}\prod_{c\in C} \psi_c(\z_c|\xibf_c),
$$
as named in statistical mechanics, ensures the probability distribution of $\z$ sums to one. 
The power of this theorem comes in that we can specify a small number of nonnegative $\psi$ functions for the cliques of the graph that imply full conditional distributions or, as noted above, verify that our specified full conditionals can be derived from Equation~\ref{eq:hc_thm} \citep[pg. 195]{rue_held2010}.

For our MRF versus MDGM comparisons, we set
$$
\psi(\z_c|\beta) = \exp(\beta I(z_i = z_j))
$$
in Equation~\ref{eq:hc_thm} for all pairwise cliques and set all other clique functions to zero.
Then, we have
\begin{equation}
\label{eq:mrf_prop_to}
    p(\z|\beta) \propto \exp\left(\beta\sum_{i\sim j} I(z_i = z_j)\right),
\end{equation}
where $i\sim j$ indicates 
the set of all neighbor pairs in the NUG, or equivalently
the edge set $\bar\E$. The full conditional of $z_i$ given all other $\z_{-i}$ is simply the conditional distribution of $z_i$ given its neighbors
\begin{equation}
\label{eq:mrf_full_cond}
p(z_i|\z_{\partial(i)},\beta) =\frac{\exp\left(\beta\sum_{j\in\partial(i)}I(z_i = z_j) \right)}{\exp\left(\beta\sum_{j\in\partial(i)}I(z_j = 0) \right) + \exp\left(\beta\sum_{j\in\partial(i)}I(z_j = 1) \right)},
\end{equation}
a version of the Ising model from statistical mechanics as it is commonly parameterized in the spatial statistics setting:  $\beta$ governs the number of matches of the values of the vertices with respect to the edges of the graph. As $\beta$ increases, the areal units will appear to be more clustered. 

An MRF prior specification for $p(\z|\thetabf)=p(\z|\xibf)$ also leads to a valid posterior distribution.
The difficulty of sampling from the posterior in an MCMC scheme comes from the partition function $R(\xibf)$, which involves the unknown parameters $\xibf$.
Methods exist to obtain valid posterior samples by avoiding evaluation of the partition function \citep{moller_etal2006, murray_etal2006}, but rely on the computationally burdensome coupling from the past algorithm of \citet{propp_wilson1996}.

\subsection{aMRF Prior}
\label{sec:pl_approx}
As another prior for comparison to the MDGM, we include the popular aMRF model (our name) where the analytical form of the pseudo-likelihood is used in place of an MRF prior to facilitate Bayesian inference via MCMC.  
In the aMRF prior, we replace the MRF distribution, $p(\z|\xibf)$, with the tractable approximation
$$
g(\z|\xibf) = \prod_{i=1}^n p(z_i|\z_{\partial(i)},\xibf),
$$
the product of the full conditionals.
For our model comparisons, we use the full conditionals of Equation~\ref{eq:mrf_full_cond} for the aMRF prior. 
\citet{besag1975} named $g(\z|\xibf)$ the pseudo-likelihood and used a factorization technique to show that the pseudo-likelihood is a consistent estimator of $\xibf$, but notably did not suggest it replace the analytical form of the MRF in a Bayesian setting.
Regardless, we often find that inference is then carried out using the approximation
\begin{equation}
\label{eq:pl_posterior_approx}
 p(\z|\xibf)p(\xibf)p(\y|\z) \approx g(\z|\xibf)p(\xibf)p(\y|\z),
\end{equation}
even though it has been noted that the aMRF, $g(\z|\xibf)$, may not correspond to a valid probability distribution \citep{friel_etal2009}, a fact we formalize in the following proposition. 

\begin{proposition}
\label{prop:amrf}
    The aMRF prior, $g(\z|\xibf)$, defined as the product of the full conditional distributions derived from an MRF, does not correspond to a valid probability distribution, when there exists a clique function for clique size $|c|>1$ such that $\psi_c(\z_c|\xibf_c)\ne t$, where $t$ is some nonnegative constant.
\end{proposition}

We provide a proof of Proposition~\ref{prop:amrf} in Appendix~\ref{ap:amrf} using the same factorization technique as in \citet{besag1975}. 
Now, as $g(\z|\xibf)$ is not a valid probability distribution, there is no guarantee (that we are aware of) that sampling from the right hand side of Equation~\ref{eq:pl_posterior_approx} (as described in Section~\ref{sec:mod_fit}) will yield draws from a valid posterior distribution, let alone the posterior in Equation~\ref{eq:general_posterior_valid}. 

\section{Model Fitting}
\label{sec:mod_fit}
We carry out posterior inference through Markov chain Monte Carlo (MCMC). 
The MCMC scheme for each spatial prior (MDGM, MRF, and aMRF) all follow the same sequence of steps.
First in the MDGM models, model fitting is facilitated by treating the DAG as a hidden state which can be updated for each $b=0,\dots,B-1$ iterations of the MCMC.
Since the NUG is fixed for the MRF and aMRF, the graph update step is skipped for these priors.
Next, we update the configuration of the latent variable, $\z$.
Third, we update the spatial dependence parameter, $\beta$, with a Metropolis-Hastings step. 
In the MRF, this MH step requires an exact sample from the MRF distribution to cleverly avoid calculating the normalizing constant \citep{moller_etal2006, murray_etal2006}.
Lastly, given the latent configuration $\z$, the update for $\etabf$ is the same for all models.
Table~\ref{tab:mcmc} gives a summary of the MCMC algorithms for our MDGM classes and for the standard MRF prior, both for exact inference and for the aMRF.
We refer the reader to Appendix~\ref{ap:mcmc} detailed executions of each step in the presented algorithms and the details for the time complexity.
Of note for the spanning tree class, the run time depends on the structure of the NUG, with denser NUGs enabling faster sampling of spanning trees from the posterior.
In the standard spatial setting, the time complexity of MCMC with MDGM-ST prior is $O(n\log n)$.
For the models with MDGM-R, MDGM-AO and aMRF priors the time complexity of the MCMC is $O(n)$, whereas MCMC for exact inference with an MRF prior has time complexity $O(n^3)$.
See Appendix~\ref{ap:time} for more details. 

\begin{table}[ht!]
    \centering
    \begin{tabular}{lp{.16\textwidth}p{.16\textwidth}p{.16\textwidth}p{.16\textwidth}p{.16\textwidth}}
        \toprule
        & \multicolumn{3}{c}{MDGM} & \multicolumn{2}{c}{} \\
           \cmidrule{2-6}
        Step & ST & Rooted & AO & aMRF & MRF \\
          \toprule
        1: $\G$ &  \multicolumn{1}{p{.16\textwidth}|}{Sample $\D \sim p(\D|\z,\beta,\y)$} &
        \multicolumn{2}{p{.32\textwidth}|}{Sample $\D^*\sim q(\D^*)$ and accept with MH ratio probability in Equation~\ref{eq:mh_ratio_graph}} &  \multicolumn{2}{p{.32\textwidth}}{$\N$ is fixed.} \\
        2: $\z$ & \multicolumn{3}{p{.48\textwidth}|}{Sample $z_i \sim p(z_i|\z_{-i},\D,\beta,\etabf,\y_i)$} &\multicolumn{2}{p{.32\textwidth}}{Sample $z_i \sim \newline p(z_i|\z_{\partial(i)},\beta,\etabf,\y_i)$} \\
        3: $\beta$ & \multicolumn{5}{p{.8\textwidth}}{Sample $\beta^* \sim q(\beta^*|\beta)$ and accept with MH ratio 
        probability}  \\ 
        &  \multicolumn{4}{p{.64\textwidth}|}{in Equation~\ref{eq:mh_ratio_beta}} & in Equation~\ref{eq:mh_ratio_beta_mrf} \\
        4: $\etabf$ & \multicolumn{5}{p{.8\textwidth}}{Sample $\eta_1 \sim p(\eta_1|\eta_0,\y,\z)$ and $\eta_0 \sim p(\eta_0|\eta_1,\y,\z)$} \\
        \bottomrule
    \end{tabular}
    \caption{MCMC schemes for a hierarchical model with an MDGM, MRF, or aMRF prior on the latent variable.}
    \label{tab:mcmc}
\end{table}

\section{Model Evaluation}
\label{sec:mod_eval}
%

As we propose the MDGM as an alternative for modeling spatial dependence, we now compare how well two classes of the MDGM perform to the classic MRF, as well as to the aMRF, a persistent default substitute for the MRF in Bayesian analyses.
For our simulation study we generate data from our model described in Section~\ref{sec:priors}, but using an MRF as the data generating mechanism for the latent variable. 
For a single simulated dataset, we generate a latent binary field, $\z$, for a specified NUG and spatial dependence parameter $\beta$.
Then, given the latent $\z$, we generate a vector of independent binary observations $\y_i$ with Bernoulli errors $\eta_0$ and $\eta_1$ at each areal unit for $i=1,\dots,n$.
Using only the observed $\y$ as input, we perform full Bayesian analysis using the MCMC algorithms described in Section~\ref{sec:mod_fit} and Appendix~\ref{ap:mcmc} for the spanning tree and acyclic orientation classes of the MDGM as well as for the MRF both without any approximations and using an aMRF.
The main goal of the simulation study is to evaluate ``how well'' each model estimates the latent spatial field given the observed binary response. 


We evaluate model performance by the posterior mean accuracy of the estimated latent field to the ground truth. 
That is, we generate the posterior distribution of accuracy by finding the rate of correspondence between each draw of the latent field and the ground truth and average over these individual estimates to obtain the posterior mean accuracy.
Additionally, we evaluate model performance by how well the different models allow us to estimate the true strength of spatial dependence assumed in the data generation process.
Using the MRF as the data generating mechanism gives us an inherent metric to evaluate spatial dependence, namely its sufficient statistic $T(\z)=\sum_{i\sim j}I(z_i = z_j)$, the number of neighboring pairs of areal units with the same value.
For a single dataset of the study, we find the posterior root mean square error (RMSE) of the sufficient statistic of the draws from the posterior to the $T(\z)$ of the generated data. 
Note, we are not interested in evaluating how well each model estimates the spatial dependence parameter, $\beta$, as the mathematical interpretation of this parameter is inherently different for each of the classes of the MDGM and for an MRF. 

Again, we emphasize that the MDGM is proposed as an \textit{alternative to}, not approximation of, an MRF; nevertheless, due to the prevalence of the MRF as a model for spatial dependence between areal units, we also include a measure of how well the posterior of each model with an MDGM or aMRF prior approximates the posterior of the model with an MRF prior when the true data generating mechanism is known to follow an MRF.
For each simulated dataset we compare the posterior distribution of $T(\z)$ for each MDGM class and the aMRF to the posterior distribution of $T(\z)$ from an MRF through the total variation distance.
The model with an MRF prior is correctly specified and can be viewed as the true posterior for the simulated data. 
We calculate the empirical total variation distance from $B$ posterior draws as
$$
\hat\delta(p(T|\y),p^*(T|\y)) = \frac{1}{2 B}\sum_{t=0}^{|\bar E|} \left|\sum_{b=1}^B I(T(\z^{b}) = t) - \sum_{b=1}^B I(T(\z^{*b}) = t)\right|,
$$
where $p(T|\y)$ is the posterior of $T(\z)$ with an MRF prior and $p^*(T|\y)$ is the posterior of $T(\z)$ with an alternate prior.
Correspondingly, $\z^{b}$ is the $b$th draw using the MRF prior and $\z^{*b}$ is the $b$th draw using the alternative prior. 

For each distinct setting in our simulation study, i.e. different values of $\beta$, $\etabf$, and $m_i$ (the number of observations at each areal unit), we generate one hundred datasets and summarize model performance by estimating the expected posterior mean accuracy, expected posterior RMSE of $\hat T(\z)$ and expected total variation distance by taking the average across the posteriors fit to the hundred simulated datasets.
For the MCMC algorithms, we initialize $\etabf^0$ at the simulation setting of $\etabf$, $\beta^0$ at the simulation setting for the MRF and aMRF priors and at $2\beta$ for the MDGM priors, and $\z^0$ at an independently and randomly generated configuration. 
We generate five thousand draws from the posterior, discarding the first one thousand as burn-in. We calculate the split potential scale reduction factor (R-hat) for the posterior draws of $\beta, \eta_0, \eta_1$ and $T(\z)$ to assess convergence of each posterior \citep{bda3}.
Posteriors that did not converge in one of the parameters or $T(\z)$ (R-hat greater than 1.1) were not included in the Monte Carlo estimate. 
The MDGM-AO and aMRF indicated a lack of convergence in the high uncertainty regimes when $\beta\in\{0.6,0.7,0.8\}$, with on average about 14 and 3 non-converged posteriors per 100 replicates, respectively.
The data generation and posterior simulation were performed in \texttt{R} \citep{r2024}.

We evaluate the MDGM models in the missing data setting, with a fixed rate of missingness across the areal units. We also include a complete data setting, where there is at least one observed $y_i$ for $i=1,\dots,n$, in Appendix~\ref{ap:complete_data}. 
We find that the latent spatial field is only partially identifiable \citep{gustafson2015}  when $1 \ge m_i \ge 0$ for $i=1,\dots,n$.
Correspondingly, in all settings of the simulation study, we have $m_i > 1$ for at least one areal unit, a constraint that improves identifiability of the $\etabf$ and $\z$.
The latent field is generated for a $16\times16$ regular lattice (256 areal units) with a first-order dependence structure, a common default.
The smaller sized lattice allows us to perform the computationally expensive exact inference for the MRF.
In Appendix~\ref{ap:sim_study}, we also evaluate the MDGM-ST and aMRF on a $32\times32$ regular lattice (1024 areal units) in the missing and complete data settings under the high uncertainty regimes.
We use the exchange algorithm to obtain an exact draw from the MRF distribution needed for the Metropolis-Hastings ratio in Equation~\ref{eq:mh_ratio_beta_mrf}.
For all simulated datasets we set $\eta_0=\eta$ and $\eta_1=1-\eta$.
First, we generate $m_i\sim\mathrm{pois}(\lambda)$. We set $\eta=0.1$ and generate one hundred datasets across each combination of $\beta=\{0.1,0.2,0.3,0.4,0.5,0.6,0.7,0.8\}$ and $\lambda=\{1.39, 2.3\}$, which corresponds to a missing rate of 25\% and 10\% respectively. 
The selection of the values for the spatial dependence parameter, $\beta$, to use in the simulation study were driven by a well studied property of the Ising model in statistical mechanics and statistics, \textit{phase transition} \citep[see][]{georgii2011}, which we explain in Appendix~\ref{ap:phase_transition}.

The simulation study was run on a Dell PowerEdge R640 server with dual 18-core Intel CPUs and 256 GB RAM, with each replicate using a single thread.
Across all datasets of the study, the average elapsed computation time of the aMRF was the fastest, with an average of 0.8 seconds to obtain five thousand draws from the posterior.
The two MDGM models had similar average elapsed times, 1.7 seconds for the spanning tree class and 1.8 seconds for the acyclic orientation class.
The exact MRF model averaged 62.3 seconds.

\begin{figure}[ht]
    \centering
    \includegraphics[width=\textwidth]{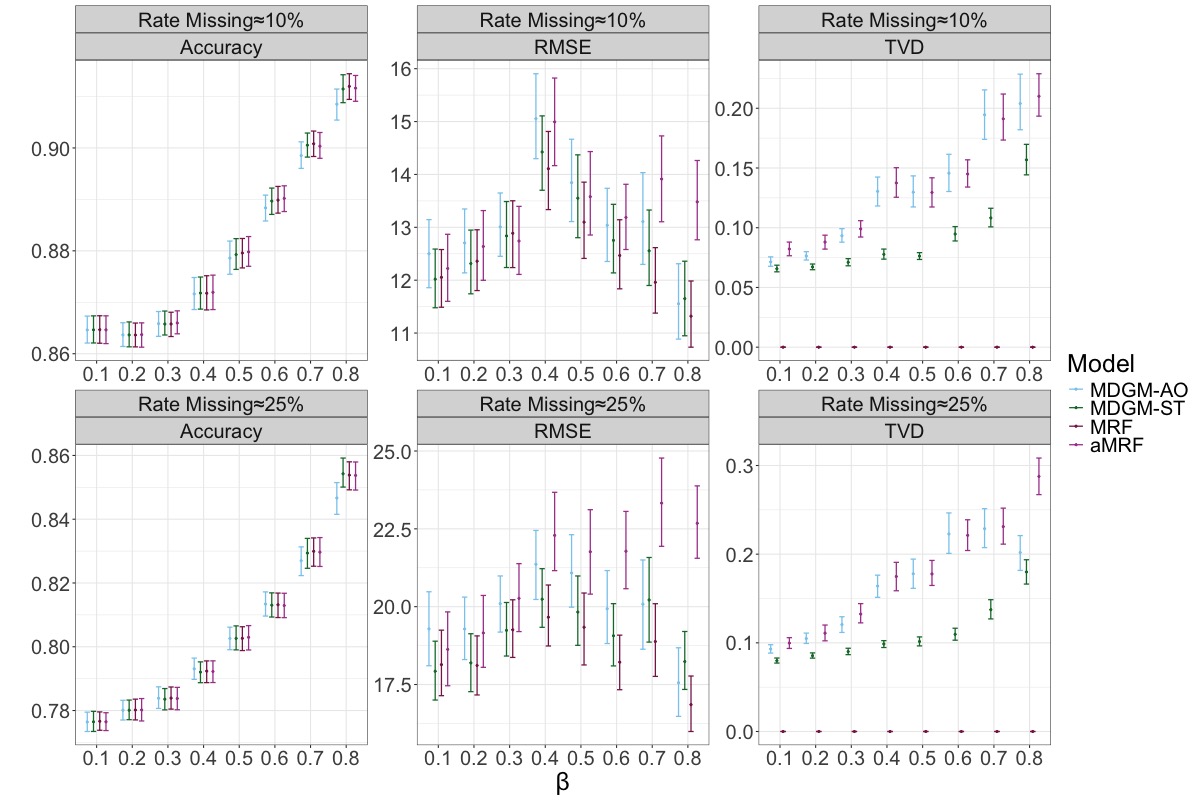}
    \caption{Simulation study results for the missing data setting. The points are the estimated expected value of the statistic with error bars giving the 90\% bootstrap confidence interval of the Monte Carlo error.}
    \label{fig:missingness}
\end{figure}

The results of the simulation study, the estimates of the expected statistic with a 90\% bootstrap confidence interval of the Monte Carlo error, are shown in Figure~\ref{fig:missingness}. 
The bootstrap confidence interval estimates were obtained by taking one thousand samples, of size one hundred and with replacement, of the expected posterior statistics across the datasets, then taking the average of each resample to estimate the sampling distribution. 
We took the fifth and ninety-fifth quantiles of the estimated sampling distribution to be our lower and upper bounds of the confidence interval. 

The three metrics each share different information on model performance. 
All models perform equally well in terms of accuracy, in both the high ($\lambda=1.39$, 25\% missing rate) and low uncertainty ($\lambda=2.3$, 10\% missing rate) settings;
The expected areal-unit-wise error rate is about the same regardless of prior.
The models do not perform equally well in capturing the true spatial dependence of the latent process.
In particular, the aMRF model has a higher expected posterior RMSE as the spatial dependence parameter increases.
This worse performance for $\beta \ge 0.5$ is more pronounced in the high uncertainty setting. 
The MDGM-ST consistently performs on par with the MRF in recovering the true spatial dependence.
While there is greater uncertainty in the Monte Carlo estimates for the MDGM-AO, this model also performs well and beats the aMRF in the high uncertainty setting when $\beta$ is large. 
Lastly, the total variation distance gives us a measure of approximation when an MRF is the true data generating distribution. 
Total variation distance ranges from zero to one. 
For all three substitute priors, the posterior probabilities over the number of matches with respect to the NUG do differ from that of the posterior for a model with an MRF prior.
In both the low and high uncertainty regime, it is clear that the MDGM-ST is a better approximation to the MRF than the MDGM-AO or aMRF priors. 
While not presented as an approximation, the MDGM-ST turns out to be a better approximation to an MRF than the commonly used aMRF. 

Across all metrics, the spanning tree class performs the best. 
As spanning trees are strongly compatible, conditionally, false dependencies not present in the NUG are not introduced.
As a result, spanning trees can come closer to approximating an MRF distribution which perfectly encodes the dependencies of the NUG.
In Appendix~\ref{ap:ci_mdag}, we explain that the MDGM cannot represent the exact same set of conditional independence relationships as the NUG due to the implicit indicator variable for the DAG structure; however,
conditioning on the DAG, in the hierarchical set up of the model, a collection of strongly compatible DAGs can encapsulate all the conditional independencies represented in the NUG.
Notably, the aMRF is a worse representation of the conditional independence relationships of the NUG because it introduces v-structures through its product mixture formulation of the sampling distribution.
See Appendix~\ref{ap:proper_colorings} for more details on how the aMRF is an unnormalized product mixture.

\section{Application to Survey Reports of Physical Disorder}
\label{sec:phys_dis_example}
In this section, we compare the MDGM-ST and the aMRF on an analysis of survey data from the Adolescent Health and Development in Context (AHDC) Study. 
This longitudinal study was designed to understand how social processes affect developmental outcomes of youth in the Columbus, Ohio metropolitan area. 
A representative sample of households residing within the I-270 belt loop and with youth ages 11-17 was obtained and a single youth from each household was randomly selected to participate in the study. 
Upon enrollment into the study, a caregiver of the youth, usually the mother, answered a questionnaire with a variety of questions about the conditions of their own neighborhood, as well as about locations which they routinely visit.
We use data from the first wave of the study, which was collected between 2014-2016. 
One subset of questions -- consistent with the ecometrics approach in urban sociology \citep{raudenbush_sampson1999} -- ask participants to report on aspects of the physical disorder of each of their routine activity locations, i.e. presence of garbage, graffiti, needles, liquor bottles, etc.
For this analysis, we analyze the respondents' perceptions of garbage, which was asked for neighborhoods as ``In your neighborhood are [garbage, litter, broken glass] a big problem, somewhat of a problem or not a problem?" and for routine locations ``Let us know if you consider [garbage, litter, broken glass] a big problem, somewhat of a problem, or not a problem at [location X]".
We recoded the three-scale Likert-type responses to a binary outcome with one corresponding to ``a big problem'' and ``somewhat of a problem'' and zero corresponding to ``not a problem''.
The spatial units of analysis are the 615 census block groups within the I-270 belt loop.
We specify a first-order dependence structure for the NUG in this analysis.
The caregiver ratings are assumed to be conditionally independent within each block group.
In total, there are 9469 ratings across the 615 block groups, of which 34 block groups have no ratings. 
Of the block groups that do have ratings, the median number of ratings is 9, the max is 192, and the third quartile is 19. 

\begin{figure}[ht]
    \begin{minipage}{.28\textwidth}
	\centering
	\includegraphics[width=.95\linewidth]{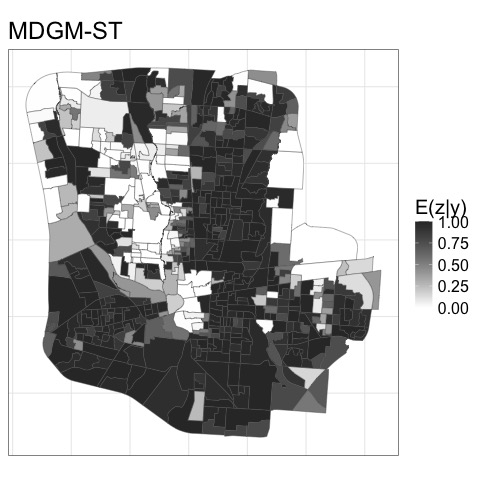}
    \end{minipage}
    \begin{minipage}{.28\textwidth}
	\centering
	\includegraphics[width=.95\linewidth]{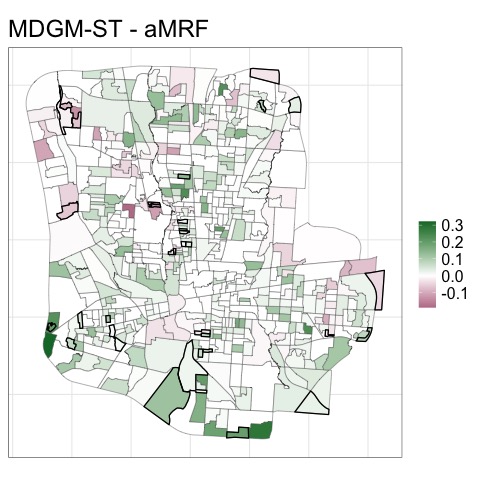}
    \end{minipage}
    \begin{minipage}{.42\textwidth}
	\centering
	\includegraphics[width=.95\linewidth]{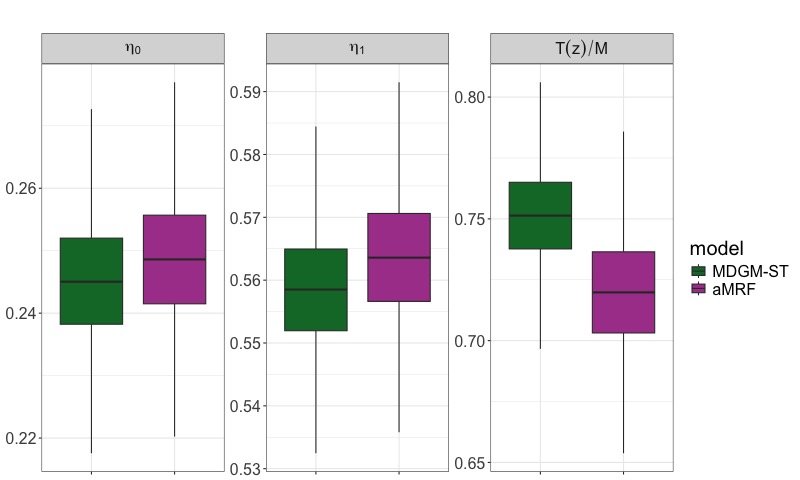}
    \end{minipage}
\caption{Left: map of the posterior mean of the latent process for each block group within the I-270 belt loop around Columbus, Ohio with a MDGM-ST prior.
Center: The difference between the block group posterior mean of the latent variable for the MDGM-ST minus the aMRF prior. The block groups outlined in a thicker black line indicate units with no ratings. Right: Comparison of the posterior distributions for $\etabf$ and $T(\z)/M$, where $M=|\bar\E|$ for the two different priors on the latent variable.}
\label{fig:disorder_comparison}
\end{figure}

Exact posterior inference for an MRF prior is already computationally infeasible for an irregular lattice with 615 areal units.  Therefore, we compare the MDGM-ST prior on the latent variable, the best performing MDGM class from the simulation studies, to an aMRF prior. 
 
We ran four MCMC chains in parallel for each model, with 5000 draws per chain, using the MCMC algorithms described in Section~\ref{sec:mod_fit} and Appendix~\ref{ap:mcmc}.
The latent variable was independently and randomly initiated. We initiated $\eta_0$ with random uniform draw on $(0,.5)$, $\eta_1$ on $(0.5,1)$, and $\beta$ on $(0,0.88)$ for the aMRF and $(0,2)$ for the MDGM-ST.
We discarded the first 1000 draws of each chain as burn-in. 
The estimated R-hat values for each parameter and the statistic $T(\z)$ were all less than 1.01 and trace plots of the four chains for each parameter showed good mixing.

Figure~\ref{fig:disorder_comparison} shows the posterior mean of the latent variable at each block group for an MDGM-ST prior in the left panel and the difference between the MDGM-ST minus the aMRF prior in the center panel. 
The difference between models is small and both produce a similar partition of the block groups into higher and lower probability areas of ratings indicating that garbage is a problem. 
Block groups with zero ratings are outlined in black.

In the right panel of Figure~\ref{fig:disorder_comparison} are box plots of the posterior distributions for the Bernoulli error parameters $\etabf$ and $T(\z)/|\bar\E|$.
Overall, the MDGM-ST estimates slightly higher probabilities of the latent variable being one compared to the aMRF, but has lower estimates of the error parameter. 
Additionally, the MDGM-ST estimates a higher proportion of matching neighboring block groups with respect to the NUG than the aMRF.
In Appendix~\ref{ap:cross_validation}, we detail our comparison the performance of the two models through a cross-validation study by holding out data in training the model and then evaluating how well each model can predict the held out areal unit ratings. 
In the cross validation study, the performance between the MDGM-ST and aMRF priors is practically equivalent.
The advantage of the MDGM-ST is the validity of the posterior distribution.
An additional advantage of the spanning tree class is the posterior edge inclusion probabilities, which provide a spatially explicit summary of local dependence learned from the data; see Appendix~\ref{ap:edge_inclusion} for details.

\section{Related Work and Discussion}
\label{sec:disc}
While we do not propose the MDGM framework as a formal approximation to an MRF model, we note that special cases of the framework appear in the literature as MRF approximation techniques. In particular, \citet{cressie_davidson1998} proposed partially-ordered Markov models (POMMs) as a DGM-based alternative to an MRF model.  \citeauthor{cressie_davidson1998} show that any POMM can be expressed equivalently as an MRF, and that for some MRFs, there is a POMM which closely approximates the probabilities of the configurations of a graph.  We note that 
the DAG associated with the POMM is in the set of DAGs compatible (as defined in Section \ref{sec:nug_as_template}) with the NUG associated with the MRF.
In this way, POMMs are a special case of the MGDM framework in which there is only one component in the mixture.
Other modeling strategies which use the computational advantages of DGMs include the Markov mesh model \citep{abend_etal1965}, the
POMM-inspired DGM approximation of the Potts model (the categorical extension of the two state Ising model) \citep{chakraborty_etal2022}, and a POMM-inspired approximation of the normalizing constant of an MRF \citep{tjelmeland_austad2012}.  

We also note the flexibility of the MDGM modeling framework.  In particular, the parent conditional distributions, which determine the form of the DGM components in 
Equation~\ref{eq:dgm_factorization}, can be any valid probability distribution (e.g., Gaussian, Poisson).
Additionally, as noted in the introduction,  the MDGM can be used directly as a model for discrete outcomes, analogously to autologistic models where an MRF is used to directly model the data \citep{besag1974, hughes_etal2011, caragea_kaiser2009}, instead of being used as a prior distribution in a hierarchical model as described in Section \ref{sec:priors}.
Lastly, the spanning tree class provides a more flexible characterization of spatial dependence through a global spatial dependence parameter and local adjustments from posterior edge weights. 

An alternative Bayesian approach for categorical spatial data uses data augmentation for probit regression \citep{albert_chib1993} with a Gaussian MRF (CAR) prior on a continuous latent variable \citep{berrett_calder2012, schliep_hoeting2015}, but this formulation suffers from partial identifiability of the spatial dependence parameter \citep{carter_etal2024}.
An intrinsic CAR specification avoids this issue but yields an improper prior, precluding prediction at unobserved locations and limiting cross-validation assessments.
The MDGM framework, by contrast, offers proper posterior predictive distributions and readily supports cross-validation model comparisons.

Another common model for discrete spatial data is the hidden Markov random field (HMRF), where the two-state latent field is expanded to multiple states and then related to a continuous outcome variable.
In this setting, the main goal of analysis is often classification; thus, full Bayesian posterior inference is often unnecessary and point estimates or variational methods are reported \citep{zhang_etal2001, freguglia_etal2020}.
Notably, \citet{moores_etal2020} do provide posterior inference in this setting, but through a surrogate model of the MRF.
Indeed, in the setting of Bayesian inference for a latent discrete-valued random field on an areal lattice, the PFAB method of \citet{moores_etal2020} is the only modern competitor with publicly available software; the ordered conditional approximation of \citet{chakraborty_etal2022} was shown by its own authors to perform worse than PFAB in recovering spatial dependence, and no software implementation is available.
In Appendix~\ref{ap:gaussian_sim}, we compare the MDGM to the PFAB method in a Gaussian emission simulation study with $k \in \{3,4,5,6\}$ classes and find that the MDGM-ST and aMRF substantially outperform PFAB across all scenarios, while the MDGM-ST remains competitive with the aMRF.
In situations where full posterior inference is desirable, the MDGM is an attractive alternative to the HMRF.

In the continuously-indexed spatial data setting, we note that our work also parallels recent developments in scalable inference for Gaussian process (GP) models, for which there has been interest in exploiting the computational advantages of DGMs to fit GP models to massive datasets \citep{vecchia1988, stein_etal2004, datta_etal2016, guinness2018, katzfuss_guinness2021, katzfuss2020, dey2022graphical}.
Recent advances include the meshed GP of \citet{peruzzi_etal2022}, which assigns a DAG to the subsets of a partition of the domain; 
a DGM approximation of Nearest Neighbor GPs to improve computation of spatial probit linear mixed models \citep{saha_etal2022}; and a ``Bag of DAGs'' model to improve computational speed and learn prevailing wind patterns in the spread of air pollutants \citep{jin_etal2023}.

In conclusion, the MDGM is a flexible and computationally efficient framework for modeling observations associated with spatial areal units. 
Most importantly for the discrete outcome setting, the MDGM offers a computationally fast and simple implementation alternative to the widely used, yet theoretically tenuous, aMRF.
Additionally, in the variety of data settings described above, the MDGM offers a flexible alternative to standard models.

Beyond applying the MDGM to different data settings (e.g., continuous outcomes, multicategory outcomes), in future work we will explore alternative MCMC schemes -- as an alternative to our independent MH proposals -- for the acyclic orientation and rooted MDGM classes. 
Through our simulation studies, we found the spanning tree class of MGDMs performed particularly well in estimating the latent field, both in terms of accuracy and capturing the spatial dependence of the field.
In terms of approximation, the strong compatibility of spanning trees provides a probable theoretical reason why the class comes closest to the posterior with an MRF prior in our simulation studies. 
The minimally connected ST-DAGs appear to offer greater flexibility in modeling the spatial dependence across areal units compared to the acyclic orientation class when the data are generated from an underlying MRF.
We will explore whether this result holds for other data generating mechanisms.

\bigskip
\noindent\textbf{AI Disclosure.} Claude (Anthropic, claude-opus-4-6) was used to assist with code organization, manuscript editing, and preparation of supplementary materials. All scientific content, methodology, and analysis were performed by the authors.

\bigskip
\begin{center}
{\large\bf SUPPLEMENTARY MATERIAL}
\end{center}

\begin{description}

\item[Appendices:] Includes all proofs, discussion of concepts and examples related to the MDGM, two additional MDGM classes, additional binary simulation study results, a Gaussian simulation study comparing to the PFAB method, and posterior edge inclusion probabilities for the Columbus application.

\item[Code and Data:] Code to reproduce the simulation studies, analysis, and cross-validation study in the paper, along with the anonymized AHDC garbage ratings data.
\if0\blind
Available at \url{https://github.com/jbcart/mdgm-supplement}. The \texttt{mdgm} R package, which implements the MDGM framework and MCMC algorithms described in this paper, is available at \url{https://github.com/jbcart/mdgm}.
\fi
\if1\blind
Available at [GitHub repository, anonymized for review]. The \texttt{mdgm} R package, which implements the MDGM framework and MCMC algorithms described in this paper, is available at [GitHub repository, anonymized for review].
\fi

\end{description}

\spacingset{1.25}
\bibliographystyle{JASA}
\bibliography{references}

\newpage
\appendix
\spacingset{1.75}

\section{NUGs and Compatibility}
\subsection{Neighborhood Structures}
\label{ap:nugs}
A \textit{first-order} neighborhood structure is defined such that the set of neighbors of areal unit $i$ are the units that have a border touching at more than just a point.
In a regular lattice, this corresponds to the areal units directly above, below, left, and right of $i$.
Figure~\ref{fig:lattice} depicts a three-by-three regular lattice with the corresponding NUG defined by a first-order dependence structure in Figure~\ref{fig:rook}.
A \textit{second-order} neighborhood structure defines neighbors to be the areal units that have common border or have corners that touch, giving a set of eight neighbors for any areal unit in a regular lattice that is not on the border.
Figure~\ref{fig:queen} shows the implied NUG of a second-order neighborhood structure for the three-by-three grid. 

\begin{figure}[ht]
\centering
\begin{subfigure}[b]{0.3\textwidth}
\centering
\begin{tikzpicture}
        \def\nx{3} \def\ny{3}
        \draw [black] (0,0) grid (\nx,\ny);
    \end{tikzpicture}
\caption{}
\label{fig:lattice}
\end{subfigure}
\begin{subfigure}[b]{0.3\textwidth}
\centering
\begin{tikzpicture}[>=stealth, thick, node distance=12mm, main/.style = {draw, circle}] 
\node[main] (1) {}; 
\node[main] (2) [below of=1] {}; 
\node[main] (3) [below of=2] {};
\node[main] (4) [right of=1] {}; 
\node[main] (5) [right of=2] {}; 
\node[main] (6) [right of=3] {};
\node[main] (7) [right of=4] {}; 
\node[main] (8) [right of=5] {}; 
\node[main] (9) [right of=6] {};
\draw (1) -- (2);
\draw (1) -- (4);
\draw (2) -- (3);
\draw (2) -- (5);
\draw (3) -- (6);
\draw (4) -- (5);
\draw (4) -- (7);
\draw (5) -- (6);
\draw (5) -- (8);
\draw (6) -- (9);
\draw (7) -- (8);
\draw (8) -- (9);
\end{tikzpicture} 
\caption{}
\label{fig:rook}
\end{subfigure}
\begin{subfigure}[b]{0.3\textwidth}
\centering
\begin{tikzpicture}[>=stealth, thick, node distance=12mm, main/.style = {draw, circle}] 
\node[main] (1) {}; 
\node[main] (2) [below of=1] {}; 
\node[main] (3) [below of=2] {};
\node[main] (4) [right of=1] {}; 
\node[main] (5) [right of=2] {}; 
\node[main] (6) [right of=3] {};
\node[main] (7) [right of=4] {}; 
\node[main] (8) [right of=5] {}; 
\node[main] (9) [right of=6] {};
\draw (1) -- (2);
\draw (1) -- (4);
\draw (1) -- (5);
\draw (2) -- (3);
\draw (2) -- (5);
\draw (2) -- (4);
\draw (2) -- (6);
\draw (3) -- (6);
\draw (3) -- (5);
\draw (4) -- (5);
\draw (4) -- (7);
\draw (4) -- (8);
\draw (5) -- (6);
\draw (5) -- (8);
\draw (5) -- (7);
\draw (5) -- (9);
\draw (6) -- (9);
\draw (6) -- (8);
\draw (7) -- (8);
\draw (8) -- (9);
\end{tikzpicture} 
\caption{}
\label{fig:queen}
\end{subfigure}
\caption{Example of a regular lattice and the NUGs  that can be used to represent dependence between the areal units of the lattice.}
\label{fig:ex_lattice}
\end{figure}

\subsection{Examples of Compatible and Incompatible DAGs}
\label{ap:compatible_example}
\begin{figure}[ht]
    \centering
    \includegraphics[width=\textwidth]{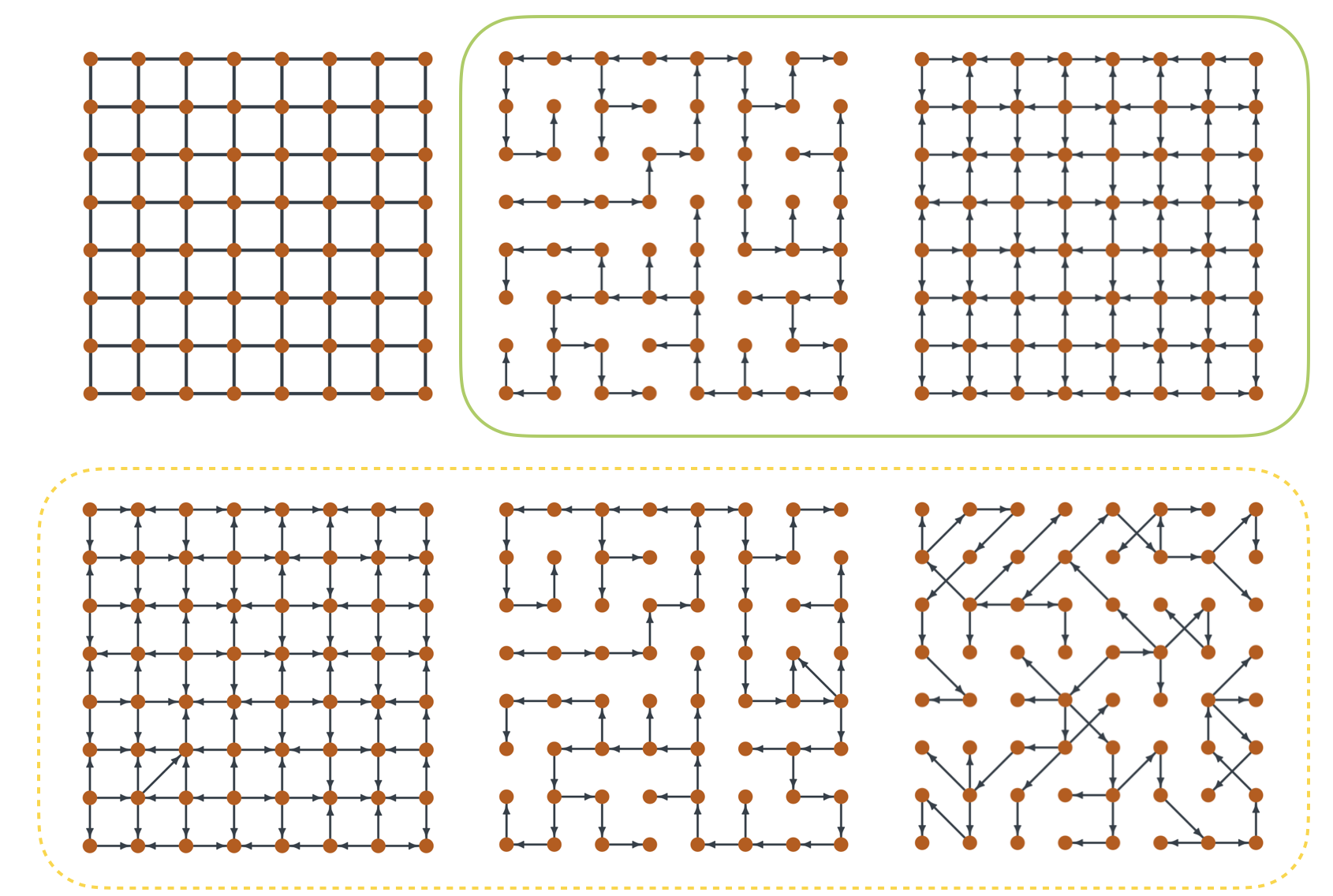}
    \caption{Example of DAGs compatible (circled by a solid line) and incompatible (circled by a dashed line) with the NUG in the top left.}
    \label{fig:compatibility}
\end{figure}

Figure~\ref{fig:compatibility} shows examples of DAGs that are compatible and incompatible with the first-order NUG shown in the top left of the figure. 
The top center DAG is an example of a directed spanning tree which is \textit{strongly} compatible with the NUG. 
The acyclic orientation DAG in the top right is \textit{weakly} compatible with the NUG. 
The bottom row shows DAGs that are incompatible with the NUG as they introduce directed edges between vertices which do not appear as undirected edges in the NUG. 

\section{Compatibility Proofs}
\label{ap:proofs}

\subsection{Proof to Theorem~\ref{thm:strongweak}}
\label{ap:proof:strongweak}

\begin{proof}
    From Lemma~\ref{lemma:subgraph} we can equivalently formulate the theorem as follows: If graph a $\G$ is a D-map of $\N$, then the skeleton of $\G$ is a subgraph of $\N$.
    For proof by contraposition, assume that $\U(\G)$ is not a subgraph of $\N$.
    Then, there exists an edge $\{i,j\}$ in the skeleton of $\G$ such that $\{i,j\}\not\in\bar\E$.
    This implies that $i\perp j|C \in I(\N)$ for some set $C$, but that $i\perp j|C \not\in I(\G)$. 
    Therefore, if $\U(\G)$ is not a subgraph of $\N$ then $I(\N) \not\subseteq I(\G)$.
\end{proof}

\subsection{Proof to Theorem~\ref{thm:spanning_trees}}
\label{ap:proof:spanning_trees}

\begin{proof}
    Let $\U$ be a spanning tree of the NUG, $\N$, and assume $A\perp B|C \in I(\N)$. Because $\U$ is a subgraph of $\N$ there exists a set $C'\subseteq C$, such that $A\perp B|C' \in I(\U)$. Therefore, it is also true that $A\perp B|C \in I(\U)$, showing that $I(\N)\subseteq I(\U)$.
\end{proof}

\subsection{Proof to Corollary~\ref{cor:st_strong_first_order}}
\label{ap:proof:st_strong_first_order}

\begin{proof}
    The proof follows by showing that any compatible DAG with more than $n-1$ edges cannot be strongly compatible with a NUG with first order dependence over a regular lattice.
    Let $\D=\{\V,\vec\E\}$ be a rooted spanning tree, compatible with a NUG, $\N=\{\V,\bar\E\}$, with first order dependence over a regular lattice.
    Assume without loss of generality that we create a new weakly compatible DAG $\D'=\{\V,\vec\E\cup(i,j)\}$ with the additional edge $(i,j)$.
    Vertex $i$ cannot be the root of the graph as this would create a directed cycle. 
    Then, by definition of a rooted spanning tree, $i$ has a single parent $k\in \partial(i)$ in $\D$, where $\partial(i)$ is defined by the NUG. 
    Because $\D'$ is weakly compatible, we also have $j\in\partial(i)$, but in a first order dependence structure $\{j,k\}\not\in\bar\E$, thereby creating an unprotected v-structure in $\D'$. 
    Therefore, $\D'$ cannot have more than $n-1$ edges and still be strongly compatible.
\end{proof}

\section{Counting and Generating Random Uniform Spanning Trees}
\label{ap:st_wilsons_size}
To generate a rooted spanning tree from a NUG, first a spanning tree is generated using the loop erased random walk algorithm, which generates uniform draws of spanning trees from a NUG \citep{wilson1996}.
A loop erased random walk proceeds by performing a random walk along the edges of a graph to generate a path of vertices that does not contain any cycles (loops). 
If the random walk returns to a vertex that has already been visited in the path, then the cycle that is created by returning to vertex, $j$, is erased from the path and the random walk then proceeds from $j$. 
To obtain a spanning tree from a NUG, randomly select a vertex of the graph to be the first vertex of the tree, by first selecting an edge with probability proportional to the edge weight and then select one of the vertices attached to that edge at random.
Then select another vertex at random by the same procedure as above and from this vertex perform a loop erased walk until the path of the random walk reaches the first vertex of the tree.
This path and the vertices thereof are now part of the tree. 
Continue to select vertices at random that are not already part of the tree and perform a loop erased random walk until the path connects to vertices that are part of the tree. 
This algorithm guarantees a uniform draw of all possible spanning trees from a NUG \citep{wilson1996}. 
The same procedure can be used to draw spanning trees where the edges have different weights, drawing spanning trees with probability proportional to the product of the weights in the tree. 
We show that the weights of the edges are informed by the latent variable in our MDGM model, providing a novel MCMC scheme to obtain draws of spanning trees from the posterior.
A rooted spanning tree is then generated by selecting a vertex to be the root, and then orienting all edges away from the root.


Kirchhoff's theorem allows us enumerate the number of spanning trees that can be generated from an NUG through submatrices of the Laplacian matrix for a graph \citep{kirchoff1847uber}.
The Laplacian of a graph is the degree matrix minus the adjacency matrix, $W(\N) - A(\N)$, where $W(\N) = \mbox{diag} \left(A(\N) \1_n\right)$ denotes the degree matrix.
By Kirchhoff's theorem, the number of spanning trees from a NUG, 
$\N$, is the cofactor of any $(i,j)$ position in the matrix.
A $(i,j)$ cofactor is calculated by deleting the $i$th row and $j$th column of the Laplacian and taking the determinant times $(-1)^{i+j}$. 

The number of rooted spanning trees from any spanning tree is the number of vertices in the graph, thus the total number of rooted spanning trees from the NUG is the number of spanning trees times the number of vertices; however, due to Markov equivalence and for reasons explained in Section~\ref{sec:priors} the effective size of the rooted spanning tree class reduces to simply the number of undirected spanning trees.


\section{Additional Classes of Compatible DAGs}
\label{ap:classes_of_dags}
In Section~\ref{sec:mdgm_st_prior}, we highlight the spanning tree class of MDGM priors. 
The strong compatibility of spanning trees makes the class a good default choice of prior/model for spatial dependence.
The MDGM, as presented, does allow for classes which contain weakly compatible DAGs.
As thoroughly explored in Appendix~\ref{ap:ci_comp}, weakly compatible DAGs can introduce dependencies between vertices/variables that are not represented in the NUG. 
The implied dependence structure from weakly compatible DAGs, may, in some cases, actually represent the prior belief of dependence between areal units. 
In these cases, the \textit{template} NUG used to derive compatible DAGs could be subgraph of a \textit{super} NUG that represents the assumed dependence structure.
In this setup, weakly compatible DAGs from template NUG could be strongly compatible with the super NUG and thus honor the assumed dependence structure.
We present two additional classes of compatible DAGs to demonstrate the flexibility of the MDGM toolkit.

First, acyclic orientations (AO) are on the opposite end of the spectrum compared to spanning trees for fully connected DAGs that can be derived from the NUG.
By connected, we mean that there exists a path between any two vertices in the graph. 
Spanning trees contain the minimum number of edges to remain fully connected, whereas acyclic orientations preserve all the edges of the NUG.
Second, while spanning trees are the only strongly compatible DAGs for first-order dependence on a regular lattice, NUGs with second-order dependence permit denser DAGs that are still strongly compatible. 
We present an algorithm for one such class, rooted DAGs, which are strongly compatible for second-order NUGs. 
While first-order dependence structures are a common default choice for the NUG, the additional classes of DAGs we present here may serve as starting points for developing new classes of DAGs, with greater density than spanning trees, that are strongly compatible with a second-order neighborhood structure. 
We review the algorithms for generating AO-DAGs from a NUG and present our algorithm for generating rooted DAGs.



\subsection{Acyclic Orientations}
\label{ap:ao}
Generally, the process of defining a directed graph from an undirected graph is known as orientation, that is, each unordered pair, $\{i,j\}$, in the edge set of the undirected graph is assigned an orientation of either $(i,j)$ or $(j,i)$ to create a directed edge set.
When the assignment of orientations to the edges of graph does not lead to any directed cycles, then the orientation is called acyclic.
The acyclic orientation class is then defined as the set of DAGs
$$
\Dcal^{\text{AO}}(\N) = \{\D=\{\V,\vec\E\}:|\vec\E| = |\bar\E|\text{ and }\D\bumpeq\N\},
$$
compatible with the NUG, $\N=\{\V,\E\}$, such that all undirected edges of the NUG appear as directed edges for every DAG in the set. 
A simple way to assign an acyclic orientation from the NUG is to permute the rows and columns of $A(\N)$ and then take the lower triangular matrix as the adjacency matrix for a DAG.
As noted above, the number of acyclic orientations of an undirected graph is fewer than number of permutations of the vertices, as multiple permutations will correspond to the same acyclic orientation. 
While the theory exists to count and enumerate the number of acyclic orientations \citep{stanley1973, squire1998}, in practice, both tasks are very computationally expensive, even for orderly graph structures such as a first-order regular lattice.
For computational expediency, we will generate random acyclic orientations by permuting the indices in our MCMC algorithm in Section~\ref{sec:mod_fit}.
We refer the reader to Appendix~\ref{ap:counting_ao} for the relevant theory to count the number of acyclic orientations of a graph.

\subsection{Rooted Graphs}
\label{ap:rooted}
We use the term rooted graph to denote a directed graph for which there is a single orphan vertex.
In an R-DAG, there is a directed path from the root vertex to every other vertex in the graph. 
While there are many rooted graphs that can be created with vertex $i$ as a root, we seek to further reduce the size of this class by algorithmically assigning directions away from the root. 
Due to the algorithmic assignment of directions, an $i$-rooted R-DAG in the class is unique, thus the size of the class is simply the number of vertices.


\begin{algorithm}
\begin{algorithmic}
    \State Begin with $i$ as the root vertex.
    \For{$j \in \V$}
    \State $l(j) \gets min_{(i-j)} \sum_{\{s,r\}\in (i-j)} w(\{s,r\}) $
    \LComment{where $(i-j)$ denotes a path from vertex $i$ to vertex $j$.}
    \EndFor
    \State Define $\vec\E$ as:
    \For{$\{r,s\}\in\bar\E$}
        \If{$l(r) < l(s)$} $(s,r)\in\vec\E$
        \ElsIf{$l(r) > l(s)$} $(r,s)\in\vec\E$
        \ElsIf{$l(r) = l(s)$} $\{r,s\} \not\in \vec\E$
        \EndIf
    \EndFor
\end{algorithmic}
\caption{Algorithm to generate an $i$-rooted DAG, $\D=\{\V,\vec\E\}$, compatible with $\N=\{\V,\bar\E\}$}
\label{alg:rooted_graph}
\end{algorithm}

The following describes Algorithm~\ref{alg:rooted_graph} to create an R-DAG for any NUG.
First, assign weights to the edges of the NUG, $w(\{i,j\})$ for $\{i,j\}\in\bar\E$, as follows: edges that represent areal units with borders touching at more than a point assign the weight one, for areal units connected by borders touching at just a point assign the weight two.
Then select a vertex, $i$, to be the root.
Label all vertices by the smallest weighted path from the root vertex. 
Orient edges from lower-valued labels to higher-valued labels.
Edges between vertices with the same label are deleted. 
When the algorithm is applied to  second-order NUGs, unprotected v-structures in the R-DAG will only occur for triplets of vertices that are all mutually connected in the NUG.
For second-order dependence, the algorithm provides a class of strongly compatible DAGs with a reduced number of directed edges (therefore, an intermediate between the acyclic orientation and spanning tree classes) and cardinality equal to the number of vertices in the NUG. 

\section{MCMC}
\label{ap:mcmc}
In this appendix, we detail the steps of each MCMC algorithm to fit a model with a spanning tree, acyclic orientation, or rooted MDGM prior, MRF and aMRF prior.
See Table~\ref{tab:mcmc} for an overview of each MCMC algorithm. 
The appendix is organized to follow the same outline as the table.
For all models, we initialize the algorithm with starting values for $\etabf^0, \beta^0,$ and $\z^0$ and initialize $\D^0$ for the MDGM models.

\subsection{Update the Graph}
This step is only for the MDGM classes. The NUG is fixed for the MRF and aMRF.
\subsubsection{Spanning Trees}
For the spanning tree class, we can sample $\D\in\Dcal^{ST}(\N)$ directly from the full conditional posterior distribution $p(\D|\z,\beta,\y)$.
We can write the posterior as
\begin{equation*}
\begin{aligned}
p(\D|\z,\beta,\y) &\propto p(\z|\D,\beta)p(\D) \\
    &= \prod_{i=1}^n \frac{\exp\left(\beta I(z_i = z_{\pi(i)}) \right)}{\exp\left(\beta I(z_{\pi(i)} = 0)\right) + \exp\left(\beta I(z_{\pi(i)} = 1) \right) } \frac{1}{|\Dcal^{ST}(\N)|},
\end{aligned}
\end{equation*}
and since $|\pi(i)| = 1$, we have 
\begin{equation}
\label{eq:sp_post_weights}
p(\pi(i)=j|\z,\beta,\y) \propto \frac{\exp\left(\beta I(z_i = z_{j}) \right)}{\exp\left(\beta I(z_{j} = 0)\right) + \exp\left(\beta I(z_{j} = 1) \right) },
\end{equation}
for $j \in \partial(i)$. 
Thus, a draw of a spanning tree from the posterior can be obtained using Wilson's algorithm as described in Section~\ref{sec:st} with the modification that the random walk occurs with edge weights given by Equation~\ref{eq:sp_post_weights}.
This modification explicitly generates a ST-DAG with root at the vertex first selected to be part of the spanning tree, yet since any rooted ST-DAG with the same tree structure are equivalent, and therefore nonidentifiable, it does not matter which vertex rooted ST-DAG we choose from a given undirected tree. 
That is to say, selecting the root vertex is not informed by the data and can be chosen uniformly at random without affecting the posterior probability of the undirected tree. 
We use the ST-DGM formulation of $p(\z|\D,\beta)$ to facilitate unified computation between the MDGM classes. 

\subsubsection{Acyclic Orientations and Rooted DAGs}
For the acyclic orientation and rooted classes, we update the DAG using an independent Metropolis-Hastings (MH) step.
Let $q(\D)$ be the proposal distribution for a new DAG.
For both the rooted and acyclic orientation classes, we set the proposal distribution to be the prior.
A uniform draw from the rooted class is obtained by randomly selecting a root vertex, then generating the $i$-rooted DAG from Algorithm~\ref{alg:rooted_graph}.
In the case of the acyclic orientation class, we obtain an acyclic orientation through permutation of the vertices to assign directions to the edges in the graph (Appendix~\ref{ap:ao}).
Given the current DAG, $\D^{b}$, and proposal, $\D^*$, set $\D^{b+1}=\D^*$ with probability
\begin{equation}
\label{eq:mh_ratio_graph}
    \min\left(1, \frac{p(\z^{b}|\D^*,\beta^{b})p(\D^*)q(\D^b)}{p(\z^{b}|\D^{b},\beta^{b})p(\D^b)q(\D^*)} \right),
\end{equation}
otherwise set $\D^{b+1}=\D^{b}$.
Since the prior is used as the proposal distribution, both $p(\D)$ and $q(\D)$ cancel out of the MH ratio.

\subsection{Update the Latent Configuration $\z$}
\subsubsection{MDGM Classes}
Given $\D^{b+1}$, $\z^b$, $\etabf^b$, and $\beta^b$, an update for $\z^{b+1}$, the configuration of the latent variable, can be performed vertex by vertex for $i=1,\dots,n$ using the full conditionals
\begin{multline*}
\label{eq:dgm_full_cond_posterior}
p(z_i|\z_{-i},\D,\beta,\etabf,\y_i) \propto \\
\frac{\exp\left(\beta\sum_{j\in\partial(i)}\I(z_i = z_j) \right) \prod_{j=1}^{m_i}\eta_{z_i}^{y_{ij}}(1-\eta_{z_i})^{1-y_{ij}}}{\prod_{k\in\kappa(i)}\left[\exp\left(\beta\sum_{j\in\pi(k)}\I(z_j = 0)\right) + \exp\left(\beta\sum_{j\in\pi(k)}\I(z_j = 1) \right)\right]},
\end{multline*}
where $z_i$ depends on the variables at the vertices $\{\pi(i),\kappa(i),\pi(\kappa(i))\}$ defined by the current DAG, $\D^{b+1}$, and the observed outcome, $\y_i = [y_{i1},\dots,y_{im_i}]$. 

\subsubsection{MRF and aMRF}
For the MRF prior (for both exact inference or the aMRF), the full conditionals to obtain an update $\z^{b+1}$ are
\begin{equation*}
\label{eq:mrf_full_cond_post}
p(z_i|\z_{\partial(i)},\beta,\etabf,\y_i)=\frac{\exp\left(\beta\sum_{j\in\partial(i)}I(z_i = z_j) \right)\prod_{j=1}^{m_i}\eta_{z_i}^{y_{ij}}(1-\eta_{z_i})^{1-y_{ij}}}{\exp\left(\beta\sum_{j\in\partial(i)}I(z_j = 0) \right) + \exp\left(\beta\sum_{j\in\partial(i)}I(z_j = 1) \right)}.
\end{equation*}

\subsection{Update the Spatial Dependence Parameter}
Next, we update the spatial dependence parameter, $\beta$, through a MH step.
Let $\beta^*$ be the proposal drawn from $q(\beta^*|\beta^b)$. 
\subsubsection{MDGM and aMRF}
For the MDGM classes and aMRF, set $\beta^{b+1}=\beta^*$ with probability 
\begin{equation}
\label{eq:mh_ratio_beta}
\min\left(1, \frac{f(\z^{b+1}|\beta^*)p(\beta^*)q(\beta^b|\beta^*)}{f(\z^{b+1}|\beta^b)p(\beta^b)q(\beta^*|\beta^b)} \right),
\end{equation}
where $f(\z^{b+1}|\beta^*)$ is a place holder for $p(\z|\D,\beta)$, the MDGM prior, and $g(\z|\beta)$ when using an aMRF prior. 

\subsubsection{MRF}
Exact inference can be carried out with the clever MH ratio of \citet{moller_etal2006} and  \citet{murray_etal2006}, which avoids computing the normalizing constant, though relies on generating an exact sample from $p(\z|\beta^*)$ using the coupling from the past algorithm \citep{propp_wilson1996}.
Generate an exact sample $\z^*$ from $p(\z|\beta^*)$ and let
$$
h(\z|\beta) = \exp\left(\beta\sum_{i\sim j} I(z_i = z_j)\right) \propto p(\z|\beta).
$$
Then, set $\beta^{b+1}=\beta^*$ with probability 
\begin{equation}
\label{eq:mh_ratio_beta_mrf}
\min\left(1, \frac{h(\z^{b+1}|\beta^*)h(\z^*|\beta^b)p(\beta^*)q(\beta^b|\beta^*)}{h(\z^{b+1}|\beta^b)h(\z^*|\beta^*)p(\beta^b)q(\beta^*|\beta^b)} \right).
\end{equation}

\subsection{Update the Noise Parameter}
The update for the noise parameter $\etabf$ is the same for all models given the latent configuration.
When $p(\eta) \equiv \mathrm{Beta}(\eta|a,b)$ then the full conditionals are both truncated beta distributions,
\begin{multline*}
p(\eta_1|\eta_0,\y,\z) \propto \\
\mathrm{Beta}\left(\eta_1 \left| a_1 + \sum_{i=1}^n\sum_{j=1}^{m_i} y_{ij}I(z_i=1),\; b_1 + \sum_{i=1}^n\sum_{j=1}^{m_i} (1-y_{ij})I(z_i=1)\right.\right) I(\eta_1>\eta_0)
\end{multline*}
and
\begin{multline*}
p(\eta_0|\eta_1,\y,\z) \propto \\
\mathrm{Beta}\left( \eta_0 \left| a_0 + \sum_{i=1}^n\sum_{j=1}^{m_i} y_{ij}I(z_i=0),\; b_0 + \sum_{i=1}^n\sum_{j=1}^{m_i} (1-y_{ij})I(z_i=0)\right.\right) I(\eta_1>\eta_0).
\end{multline*}

\subsection{Time Complexity}
\label{ap:time}
We note the time complexity of each algorithm listed above in terms of the number of areal units $n$. 
For all spatial priors, steps 2 and 4 have linear time complexity.
For the MDGM priors and aMRF, step 3 is also linear.
Accordingly, the aMRF is the fastest of the three priors and has $O(n)$ time complexity.

The time complexity of the MCMC algorithm for the MDGM priors is determined by step 1, sampling a graph for a particular class.
For the acyclic orientation class, we generate orientations by sampling a random permutation of the vertices which scales linearly in $n$. 
The algorithm to generate a rooted DAG also has $O(n)$ time complexity.

For the spanning tree class, the time complexity depends on the structure of the NUG.
The run time of  Wilson's algorithm is proportional to the \textit{mean hitting time} of the graph.
The \textit{hitting time}, $H(i,j)$, is the expected number of steps that a random walk takes to arrive at vertex $j$ when starting at vertex $i$.
The mean hitting time is the hitting time averaged over all pairs of vertices in the graph,
$$
\tau = \sum_{i,j} p(i)p(j)H(i,j),
$$
where $p(\cdot)$ is the stationary distribution of the random walk. 
We can reduce the above formulation to
$$
\tau = \sum_{j} p(j)H(i,j),
$$
as the value of the sum is invariant of the starting vertex $i$ \citep{lovasz1993random}. 

To calculate the mean hitting time, we introduce a spectral representation of the NUG from \citet{lovasz1993random}.
Let $A(\N)$, be the adjacency matrix as defined in Section~\ref{sec:weak_compatibility}.
Let $D$ be a diagonal matrix with $1/|\partial(i)|$ on the $i$th diagonal. 
We define $M=D^{1/2}A(\N)D^{1/2}$, which can be represented in spectral form as
$$
M = \sum_{i=1}^n \lambda_i \vv_i\vv_i',
$$
where $\lambda_i$ and $\vv_i$ are the $i$th eigen value and eigen vector, respectively.
Then, we can calculate the mean hitting time as
\begin{equation}
\label{eq:mht_spectra}
\tau = \sum_{j} p(j)H(i,j) = \sum_{i=2}^n \frac{1}{1-\lambda_i}.
\end{equation}

While we can calculate the mean hitting time for a particular graph, the time complexity of the mean hitting time is not always available.
Another useful fact for Wilson's algorithm is that the mean hitting time is always less than the \textit{cover time} of a graph.
The cover time of a graph is the expected number of steps that a random walk takes before it visits every vertex in the graph.
The cover time is an upper bound on the maximum hitting time ($\underset{i,j}\max H(i,j)$) and thus, also an upper bound on the mean hitting time.
We list known results for the cover time and mean hitting times of common graphs. 
While we present our model in the spatial setting, which usually assumes a first or second-order neighborhood structure over areal units, our framework can be applied to a variety of modeling scenarios where a different NUG structure is used, such as a $k$ nearest neighbors NUG.

\begin{table}[ht!]
    \centering
    \begin{tabular}{l|cc}
        Graph  & Cover Time & Mean Hitting Time \\
        \hline
        Any Graph & $O(n^3)$ & $O(n^3)$ \\
        Complete & $\Theta(n\log n)$ & $\Theta(n)$ \\
        $d$-regular & $\Theta(n^2)$ &  \\
        Planar & $\Theta(n^2)$ &  \\
        2-d lattice & $\Theta(n\log^2 n)$ & $\Theta(n\log n)$ \\
        3-d lattice & $\Theta(n\log n)$ & $\Theta(n)$ \\
    \end{tabular}
    \caption{Cover and mean hitting times for common graphs}
    \label{tab:graph_times}
\end{table}

\citet{lovasz1993random} provides a nice review of random walks on graphs.
For any graph, the cover time for any graph is at least $\Omega(n\log n)$ and at most $O(n^3)$.
This upper bound is attained  by specially constructed graphs and is uncommon for most graphs.
One such pathological example is a lollipop graph where $n/2$ vertices form a clique and the other $n/2$ are a chain attached to the clique.
Notably, the mean hitting time of the lollipop graph, $\Theta(n^2)$, is substantially less than the cover time.
In general, graphs with higher connectivity have lower cover times and thereby lower mean hitting times.
For \textit{complete graphs}, where there is an edge between every pair of vertices, the cover time is $\Theta(n\log n)$ and the mean hitting time is $\Theta(n)$ \citep{aldous1983time}.
A $d$-regular graph is a graph for which every vertex has $d$ neighbors. 
The cover time of a $d$-regular graph is $\Theta(n^2)$ and the mean hitting time is $O(n^2)$. 
This bound can be improved when $d$ is large.

For a general planar graph, a graph for which the vertices can be arranged on a plane so that no two edges overlap, the cover time is $\Theta(n^2)$ \citep{jonasson2000cover}.
A first-order dependence structure over any set of areal units is planar.
The NUG for a 2-dimensional, regular lattice has cover time $\Theta(n\log^2n)$ and the mean hitting time $\Theta(n\log n)$.
Interestingly, the cover time for a 3-d regular lattice has cover time $\Theta(n\log n)$ and mean hitting time $\Theta(n)$.
The higher connectivity of second-order dependence lowers the constant of the above mean hitting and cover times for NUGs over areal units or 3-d units.
Table~\ref{tab:graph_times} provides a summary of these results.

Note that these results are for unweighted graphs, while in the MCMC algorithm we sample spanning trees from the posterior which has weighted edges. 
When the effective weight of many edges is zero, decreasing the connectivity of the graph, the sampling of spanning trees can slow down.
In our simulation studies, we did not see a slowing down, nor an increase in the variance of the run time, as the spatial dependence parameter increased.

The MCMC algorithm for exact inference for the model with an MRF prior is the slowest of the three due to the computationally intensive coupling from the past algorithm.
Coupling from the past is proportional to the mixing time of a Gibbs sampler for an MRF.
\citet{wilson1996} note a $O(n^3)$ time complexity of the algorithm, however, the mixing time is known to experience a critical slowing down as $\beta$ approaches the critical value, which could lead to exponential time complexity.

\subsection{Gaussian Emission Model}
\label{ap:mcmc_gaussian}
The MCMC algorithms described above extend naturally to Gaussian emissions and a $k$-state Potts model for the latent field ($z_i \in \{0,\dots,k-1\}$).
The hierarchical model takes the same form as Equation~\ref{eq:general_posterior_valid}, with emission parameters $\thetabf = (\mubf, \sigmabf^2) = (\mu_0,\dots,\mu_{k-1}, \sigma^2_0,\dots,\sigma^2_{k-1})$ replacing $\etabf$.
Each observation is generated as $y_i | z_i = c \sim \norm(\mu_c, \sigma^2_c)$ independently.

The four steps of the MCMC (Table~\ref{tab:mcmc}) carry over with the following modifications.

\subsubsection*{Step 1: Update the Graph}
No changes are required. For the spanning tree class, the posterior edge weights in Equation~\ref{eq:sp_post_weights} generalize to a $k$-state normalizing constant in the denominator:
$$
p(\pi(i)=j|\z,\beta,\y) \propto \frac{\exp\left(\beta I(z_i = z_{j}) \right)}{\sum_{c=0}^{k-1}\exp\left(\beta I(z_{j} = c) \right)}.
$$
As $I(z_j = c) = 1$ for exactly one value of $c$ and zero otherwise, this simplifies to
$$
p(\pi(i)=j|\z,\beta,\y) \propto \frac{\exp\left(\beta I(z_i = z_{j}) \right)}{\exp(\beta) + (k-1)}.
$$
Wilson's algorithm proceeds as before with these updated edge weights.
For the acyclic orientation class, the MH ratio in Equation~\ref{eq:mh_ratio_graph} uses the $k$-state form of $p(\z|\D,\beta)$ without further modification.

\subsubsection*{Step 2: Update the Latent Configuration $\z$}
The full conditional for each vertex $z_i$ now includes the Gaussian likelihood in place of the Bernoulli likelihood. For the MDGM classes,
\begin{multline*}
p(z_i = c|\z_{-i},\D,\beta,\thetabf,y_i) \propto \\
\frac{\exp\left(\beta\sum_{j\in\partial(i)}I(c = z_j) \right) \cdot \phi(y_i; \mu_c, \sigma^2_c)}{\prod_{l\in\kappa(i)}\sum_{c'=0}^{k-1}\exp\left(\beta\sum_{j\in\pi(l)}I(z_j = c') \right)},
\end{multline*}
where $\phi(y_i; \mu_c, \sigma^2_c) = (2\pi\sigma^2_c)^{-1/2}\exp\left(-\frac{(y_i - \mu_c)^2}{2\sigma^2_c}\right)$ is the Gaussian density.
For the MRF and aMRF priors, the analogous full conditional is
$$
p(z_i = c|\z_{\partial(i)},\beta,\thetabf,y_i) \propto \exp\left(\beta\sum_{j\in\partial(i)}I(c = z_j) \right) \cdot \phi(y_i; \mu_c, \sigma^2_c).
$$
In both cases, $z_i$ is sampled from the discrete distribution over $c \in \{0,\dots,k-1\}$ obtained by normalizing the above expressions.

\subsubsection*{Step 3: Update the Spatial Dependence Parameter}
No structural changes are needed; the MH ratio in Equation~\ref{eq:mh_ratio_beta} is used as before with the $k$-state form of $p(\z|\D,\beta)$ or $g(\z|\beta)$.

\subsubsection*{Step 4: Update the Emission Parameters}
In the Gaussian case, we place independent conjugate priors on the class-specific means and variances: $\mu_c \sim \norm(\mu_0, \sigma^2_0)$ and $\sigma^2_c \sim \mathrm{InvGamma}(\alpha_0, \beta_0)$.
To resolve the label switching problem, we impose the identifiability constraint $\mu_0 < \mu_1 < \cdots < \mu_{k-1}$.
The full conditional for $\sigma^2_c$ is
$$
\sigma^2_c | \z, \y, \mu_c \sim \mathrm{InvGamma}\left(\alpha_0 + \frac{n_c}{2},\; \beta_0 + \frac{1}{2}\sum_{i:z_i=c}(y_i - \mu_c)^2\right),
$$
where $n_c = |\{i : z_i = c\}|$ is the number of vertices assigned to class $c$.
The full conditional for $\mu_c$, subject to the ordering constraint, is a truncated normal distribution,
$$
\mu_c | \z, \y, \sigma^2_c \sim \norm(\mu_c^*, {\sigma_c^*}^2) \cdot I(\mu_{c-1} < \mu_c < \mu_{c+1}),
$$
where
$$
{\sigma_c^*}^2 = \left(\frac{1}{\sigma^2_0} + \frac{n_c}{\sigma^2_c}\right)^{-1}, \quad \mu_c^* = {\sigma_c^*}^2\left(\frac{\mu_0}{\sigma^2_0} + \frac{\sum_{i:z_i=c} y_i}{\sigma^2_c}\right),
$$
with the convention $\mu_{-1} = -\infty$ and $\mu_k = \infty$. The variance is updated before the mean in each sweep to condition on the current value of $\sigma^2_c$.

\section{Proper Colorings of Graphs}
\label{ap:proper_colorings}
The concept of a \textit{proper coloring} from graph theory is fundamental to the proof of Proposition~\ref{prop:amrf} and counting the number of acyclic orientations that can be created from a NUG.
A proper coloring of a NUG, $\N$, is an assignment of labels (i.e. categorical values) to each vertex in the graph so that no two adjacent vertices in the graph have the same label. 
The chromatic number, $\omega(\N)$, of a NUG is the minimum number of colors (i.e. distinct labels) to ensure a proper coloring. 
Let $q$ denote the size of largest clique of a graph, $\N$, then $q$ is a lower bound for $\omega(\N)$, that is $\omega(\N)\geq q$.

\subsection{Proof of Proposition~\ref{prop:amrf}}
\label{ap:amrf}
A proper coloring of a graph produces the factorization technique of \citet{besag1975}. 
The labels of a proper coloring partition the vertices into independent sets, conditional on all other vertices.
That is, the vertices of the same color are independent of all other vertices of that same color given all vertices of different colors.
Consider a NUG with chromatic number $q$, then we can factor 
$$
g(\z|\xibf) = \prod_{l=1}^q p(\z_l|\z_{-l},\xibf),
$$
where $\z_l$ is a vector of the vertices of label $l$ and $\z_{-l}$ is a vector of the vertices not of label $l$.
With this factorization, \citet{besag1975} showed that maximum pseudo-likelihood is a consistent estimator of $\xibf$ because it is a weighted average of the consistent estimators, $p(\z_l|\z_{-l},\xibf)$ (when maximizing the log pseudo-likelihood). 
Now we present our proof with the same factorization technique. 

\begin{proof}
Assume, for the sake of contradiction, that the aMRF does correspond to a proper probability distribution. 
By the condition in the proposition, there exists a clique function $\psi_c(\z_c|\xibf_c)\ne t$, where $t$ is some nonnegative constant, for $|c|>1$.
Without loss of generality, consider the case of a NUG with first-order dependence structure on a regular lattice.
The first-order NUG has a chromatic number of two, thus 
$
g(\z|\xibf) = p(\z_1|\z_2,\xibf)p(\z_2|\z_1,\xibf).
$
Because of the non-constant, non-singleton clique function,  $\z_1$ and $\z_2$ are not independent, so the above factorization shows that $g(\z|\xibf)$ will not sum to one over $\z\in\Zcal$. 
Therefore, the aMRF is not a proper probability distribution. 

\end{proof}

If all $\psi_c$ functions for clique sizes greater than one are equal to some constant, then the function $g$ does specify a valid probability distribution, a product of independent distributions -- the antithesis of a spatial analysis setting. 
The normalizing constant to ensure the aMRF does indeed sum to one is 
$
\sum_{\z\in\Zcal} g(\z|\xibf),
$
which returns us to an intractable normalizing constant that we sought to avoid in the MRF. 
Interestingly, the factorization of the aMRF given by a proper coloring of the graph is, in fact, a product mixture of DGMs. 
Each $p(\z_l|\z_{-l},\xibf)$ is a DGM represented by a DAG where all edges are oriented to the vertices of color $l$, the vertices not of color $l$ are orphans and edges between vertices not of color $l$ are deleted. 
Rather than try to find the normalization constant of a product mixture, we instead fit a standard (additive) mixture model and expand the class of possible DGMs in the mixture. 

\subsection{Counting Acyclic Orientations}
\label{ap:counting_ao}
Proper colorings also aid in counting number of acyclic orientations that can be generated from a NUG.
We can extend the concept of a proper coloring to the chromatic polynomial of a NUG, denoted $\chi(\N,p)$, which is defined as the number of ways to obtain a proper coloring of the graph using $p\in\mathbb{C}$ colors, where $\mathbb{C}$ is the set of complex numbers.
The chromatic polynomial, $\chi(\N,p)$, is equal to zero for $0\leq p<\omega(\N)$.  
\citet{stanley1973} showed that the number of acyclic orientations of a graph can be calculated as $(-1)^{n}\chi(\N,-1)$, where $n=|\V|$.
In general it is difficult to calculate the chromatic polynomial for a given graph, though results are known for special cases such as lattice strips \citep{chang2001, rocek_etal1998}.

\section{Conditional Independence Relationships in Graphs and Probability Distributions}
\label{ap:ci}
In this section, we provide the relevant background information on the expressive capabilities of probability distributions whose conditional independence relations can be represented by a graph, namely MRFs and DGMs.
Conditional independence relationships, i.e. statements of the form $\z_A\perp\z_B|\z_C$ for sets of variables indexed by $A$, $B$ and $C$, from a probability distribution are determined by the ability to factorize the joint probability distribution into independent distributions conditional on the separating set.
When $A\perp B|C$/$\z_A\perp\z_B|\z_C$, $C$/$\z_C$ is called the separating set and the joint probability distribution can be factorized as
$$
p(\z_A,\z_B,\z_C) = p(\z_A|\z_C)p(\z_B|\z_C)p(\z_C).
$$
Let $I(p)$ be the set of all conditional independence relationships asserted by the probability distribution, $p$.
A main appeal of MRFs and DGMs is the ability to use their graphical representations, NUGs and DAGs respectively, to summarize the set of conditional independence relationships of the probability distribution.
Before explaining how the graphs can be used to summarize conditional independence relationships between variables, we formally relate the probability distributions to the graphical representations through the sets of conditional independence assertions.
Let $I(\G)$ represent the set of all conditional independence relationships of the form $A\perp B|C$ asserted by the graph, $\G$.
We say that $\G$ is an independence map (I-map) of $p$ when $I(\G)\subseteq I(p)$ \citep{pearl1988probabilistic}.
In the case that the sets of conditional independence assertions are equivalent, i.e. $I(\G) = I(p)$, then $\G$ is a perfect map (P-map) of $p$. 

\begin{figure}[ht]
\centering
\begin{tikzpicture}[thick, node distance=20mm, main/.style = {draw, circle}] 
\node[main] (1) {$a$}; 
\node[main] (3) [right of=1] {$c$};
\node[main] (2) [below of=3] {$b$}; 
\node[main] (4) [right of=2] {$d$};
\node[main] (5) [right of=3] {$e$};
\node[main] (6) [right of=4] {$f$};
\draw (1) -- (2);
\draw (1) -- (3);
\draw (4) -- (5);
\draw (2) -- (4);
\draw (6) -- (4);
\draw (2) -- (5);
\draw (3) -- (5);
\end{tikzpicture} 
\caption{An example of an undirected graph.}
\label{fig:undirected_graph}
\end{figure}

As stated in the introduction, an MRF is a probability distribution whose conditional distributions have the same dependence structure as the NUG. 
In terms of conditional independence relationship sets, we can alternatively define an MRF, $p(\z)$, specified for a NUG, $\N$, as a probability distribution such that $I(p) =  I(\N)$ is true.
We previously used the local Markov property, that is a vertex, $i$, is independent of all other vertices given its neighbors, to obtain the full conditionals of $z_i$ given the set of neighbors $\z_{\partial(i)}$ (Equation~\ref{eq:mrf_full_cond}).
Additional conditional independence relationships are also asserted by the NUG. 
The most general rule to determine whether $A$ and $B$ are independent given $C$ is if all paths from $A$ to $B$ go through the set $C$.
This is known as the global Markov property.
From this definition of independence -- defined through the graphical representation of dependence between variables -- follows the local Markov property and an additional property known as pairwise Markov property. 
The pairwise Markov property states that any two nodes $i$ and $j$ are independent given all other vertices if $i$ and $j$ are not neighbors. In Figure~\ref{fig:undirected_graph}, $a\perp e | \{c,b,d,f\}$, as there is not an edge between $a$ and $e$ and all paths connecting $a$ and $e$ go through the conditioning set. 
Again in Figure~\ref{fig:undirected_graph}, we can verify that $b\perp\{c,f\}|\{a,d,e\}$ as there is no path from $b$ to $c$ or $b$ to $f$ that does not go through $a,d$ or $e$, the neighbors of $b$. 
We can prove the pairwise and local properties from the global property and, incidentally, we can use any one of the properties to prove the other two when the probability distribution over the state space is strictly positive, highlighting the intuitive power of the graphical representation \citep[pg 119]{koller_friedman2009}.

\begin{figure}[ht]
\centering
\begin{tikzpicture}[thick, >=stealth, node distance=20mm, main/.style = {draw, circle}] 
\node[main] (1) {$a$}; 
\node[main] (3) [right of=1] {$c$};
\node[main] (2) [below of=3] {$b$}; 
\node[main] (4) [right of=2] {$d$};
\node[main] (5) [right of=3] {$e$};
\node[main] (6) [right of=4] {$f$};
\draw[->] (1) -- (2);
\draw[->] (1) -- (3);
\draw[<-] (4) -- (5);
\draw[->] (2) -- (4);
\draw[<-] (6) -- (4);
\draw[<-] (2) -- (5);
\draw[->] (3) -- (5);
\end{tikzpicture} 
\caption{An example of an directed acyclic graph.}
\label{fig:dag}
\end{figure}

Now we address the conditional independence relationships expressed by a DAG.
As shown in Equation~\ref{eq:dgm_factorization}, a probability distribution $p(\z)$ defined for a graph $\D$ can be factorized into the product of the conditional distributions of $z_i$ given parents of $i$, when $p(z_i|z_{\pi(i)})=p(z_i)$ is also defined for orphan vertices.
When this factorization is unique, then $\D$ is a P-map for $p(\z)$.
The directed edges no longer represent a symmetric relationship between vertices as in the NUG, but rather that vertex $i$ depends on the set of its parents $\pi(i)$ 
Consider the DAG in Figure~\ref{fig:dag}, which was created from the graph in Figure~\ref{fig:undirected_graph}, by applying the acyclic orientation implied by the sequence, $(a,c,e,b,d,f)$.
In Figure~\ref{fig:dag}, vertex $b$ depends on $a$ and $e$, while $e$ only depends on $c$. 
We can further extend the notion of parent and children vertices to ancestor and descendant vertices. For any vertex, $j$, \textit{from} which there is a directed path \textit{to} $i$, we say $j$ is an ancestor of $i$.
Likewise, any vertex, $k$, \textit{to} which there is a directed path \textit{from} $i$, is a descendant of $i$. 
The directed and acyclic edges of a DAG imply a topological ordering of the vertices, such that if there is a directed path from $j$ to $i$ in the DAG then $j$ comes before $i$ in the ordering.
The DAG asserts a different type of Markov property that a vertex $i$ is independent of all ancestors in the graph given its parents.
The Markov property for DAGs is also known as the memoryless property as the topological ordering can be thought of as a temporal sequence: the preceding vertices of vertex $i$ are past events which don't affect $i$ given the immediately preceding events, i.e. the parents of $i$.
In Figure~\ref{fig:dag}, $d$ is independent of $a$ and $c$ given its parents $b$ and $e$.
More generally, the Markov condition guarantees that any probability distribution which factorizes according to a DAG $\D$ satisfies $I(\D)\subseteq I(p)$; that is, $\D$ is an I-map of $p$ \citep{pearl1988probabilistic, lauritzen1996graphical}.
The converse --- that $I(p)\subseteq I(\D)$, so that every conditional independence in $p$ corresponds to a d-separation in $\D$ --- is the faithfulness assumption \citep[Chapter 2]{spirtes_etal2000}.
Faithfulness is not assumed in this paper; the theoretical results in Sections~\ref{sec:weak_compatibility} and \ref{sec:strong_compatibility} rely only on the Markov condition direction, which holds by construction for distributions that factorize according to a DAG.

While the Markov property for DAGs is intuitive and leads to the convenient factorization of the corresponding DGM, more general statements about the set of conditional independence relationships in a DAG, $I(\D)$, require the concept of directed separation (d-separation). 
Recall that a path is a sequence of undirected and connected edges. 
A path in a DAG is then a sequence of connected edges where the direction is ignored. 
There are two rules that govern d-separation for sets of variables $A$ and $B$ given $C$, that is conditional independence statements of the form $A\perp B|C$. 
We say that $A$ and $B$ are separated by $C$ if all the paths from $A$ to $B$ together imply independence when conditioned on $C$.
The rules of d-separation depend on the notion of a collider.
A collider vertex at which two directed edges meet head to head. 
In Figure~\ref{fig:dag} the vertices $b$ and $d$ are both colliders.
The set of vertices that have directed edges oriented towards the collider is called a v-structure. 
$a\rightarrow b \leftarrow e$, is a v-structure in Figure~\ref{fig:dag}.
For a path that contains a collider, the path separates two sets of variables $A$ and $B$, if the collider or any of the descendants of the collider \textit{are not} in the set $C$.
Second, $A$ and $B$ are separated by $C$ if every path that does not contain a collider has a vertex that is in $C$. 
Using these two rules we can see that $a\perp e|c$, as the path $a-c-e$ does not contain a collider and we condition on $c$, whereas the path $a-b-e$ does contain a collider and $b$ is not conditioned on. 
We can also assert that $a\not\perp e |c,b$, since the path with the collider $b$ is now in the conditioning set. 
Additionally, $a\not\perp e |c,d$ and $a\not\perp e |c,f$, since $d$ and $f$ are descendants of the collider $b$. 


\subsection{Comparison of Conditional Independence Relationships}
\label{ap:ci_comp}
The following discussion is modified from the example used in Chapter 19 of \citet{murphy2012} as we further explore the differences between $I(\D)$ and $I(\N)$ and
consider the common case where $I(\D)\neq I(\N)$, nor is $I(\D)\not\subset I(\N)$, nor $I(\N)\not\subset I(\D)$.
That is, a NUG and DAG cannot be used to represent the same probability distribution, neither perfectly nor as an I-map. 
Consider the example of the NUG in Figure~\ref{fig:graph_a} which asserts that $b\perp d | a,c$ but $b\not\perp d$. 
We can create a DAG from Figure~\ref{fig:graph_a} by assigning an acyclic orientation to the vertices.
In doing so, we will invariably change the set of conditional independence relationships that are asserted by the original undirected graph. 
Figures~\ref{fig:graph_b} and \ref{fig:graph_c} are two failed attempts to construct a DAG with the same independence relationships as the undirected graph in Figure~\ref{fig:graph_a}.
In Figure~\ref{fig:graph_b}, $b\perp d$ but $b\not\perp d|a,c$, the complement of the two independence relationships we stated about Figure~\ref{fig:graph_a}.
Figure~\ref{fig:graph_c} correctly asserts that $b\not\perp d$, however, shows that $b\not\perp d|a,c$ which is incorrect for Figure~\ref{fig:graph_a}.
In fact, there is no DAG that can encode the same conditional independence relationships as the given NUG. 
Any DAG that we create from the NUG by assigning a direction to each edge in the graph will invariably result in a v-structure, which introduces dependence between previously unconnected vertices when the collider is in the conditioning set.
The introduction of v-structures is the most obvious violation of the pairwise Markov property when creating a DAG from a NUG.

\begin{figure}[ht]
\centering
\begin{subfigure}[b]{0.3\textwidth}
\centering
\begin{tikzpicture}[thick, node distance=15mm, main/.style = {draw, circle}] 
\node[main] (1) {$b$}; 
\node[main] (2) [below right of=1] {$c$}; 
\node[main] (3) [above right of=1] {$a$};
\node[main] (4) [above right of=2] {$d$};
\draw (1) -- (3);
\draw (1) -- (2);
\draw (2) -- (4);
\draw (4) -- (3);
\end{tikzpicture} 
\caption{}
\label{fig:graph_a}
\end{subfigure}
\begin{subfigure}[b]{0.3\textwidth}
\centering
\begin{tikzpicture}[>=stealth, thick, node distance=15mm, main/.style = {draw, circle}] 
\node[main] (1) {$b$}; 
\node[main] (2) [below right of=1] {$c$}; 
\node[main] (3) [above right of=1] {$a$};
\node[main] (4) [above right of=2] {$d$};
\draw[->] (1) -- (3);
\draw[->] (1) -- (2);
\draw[<-] (2) -- (4);
\draw[->] (4) -- (3);
\end{tikzpicture} 
\caption{}
\label{fig:graph_b}
\end{subfigure}
\begin{subfigure}[b]{0.3\textwidth}
\centering
\begin{tikzpicture}[>=stealth, thick, node distance=15mm, main/.style = {draw, circle}] 
\node[main] (1) {$b$}; 
\node[main] (2) [below right of=1] {$c$}; 
\node[main] (3) [above right of=1] {$a$};
\node[main] (4) [above right of=2] {$d$};
\draw[->] (1) -- (3);
\draw[->] (1) -- (2);
\draw[->] (2) -- (4);
\draw[->] (4) -- (3);
\end{tikzpicture} 
\caption{}
\label{fig:graph_c}
\end{subfigure}
\caption{An example of an undirected graph and two DAGs which do not represent the same coniditonal independence relationships between the vertices. Example modified from \citep{koller_friedman2009}.}
\label{fig:ex_dag}
\end{figure}

There are special cases, however, where $I(\D)=I(\N)$, when the skeleton of $\D$ is also equivalent to $\N$.
The skeleton of a DAG $\D$, is the undirected graph created as 
$\U(\D)=\{\V,\{\{i,j\}: (i,j)\in \vec\E\}\}$.
When two graphs, $\G$ and $\G'$, represent the same conditional independence relationships, $I(\G)=I(\G)$, they are said to be independence equivalent (I-equivalent). 
Chordal graphs are a special case of graph for which a DAG and an undirected graph can represent the equivalent sets of conditional independence relations. 
A chordal graph is a graph in which every undirected cycle of four or more vertices has edges which are not part of the cycle but connect vertices within the cycle.
The graph in Figure~\ref{fig:graph_a} is nonchordal, as there are no edges connecting the four vertices of the cycle that are not part of the cycle.
Figure~\ref{fig:graph_a} could be made into chordal graph by adding an edge from either $a$ to $c$, $b$ to $d$, or both. 
Any DAG created from a chordal NUG, $\G$, by assigning a direction to each edge in $\G$ and does not introduce v-structures for vertices whose parents are not already connected, will represent an equivalent set of independence relationships as asserted by $\G$.
By definition of chordality, in chordal graphs it is always possible to find an acyclic orientation in which the parents of a collider, as defined by the acyclic orientation, are already connected.
The NUG, $\N$, is said to be an equivalence class of all DAGs generated by assigning an acyclic orientations to $\N$ which do not introduce v-structures for unconnected parents \citep{he_etal2015}. 
While practically zero areal data analysis problems admit a chordal graph to define the neighborhood structure between areal units, as referenced in Section~\ref{sec:mdgm}, we utilized this relationship to our advantage when subgraphs of the NUG are chordal in order to create DAGs which are strongly compatible with the NUG.
We also used this relationship to our computational advantage for tree graphs, another special instance of a chordal graph.
Since a tree is a chordal graph, the undirected graph is an equivalence class for all DAGs with the same skeleton in which all edges are oriented away from a single vertex, called the root.
Again, it can be shown that specifying a probability distribution for a tree graph can be written equivalently as an MRF or as a DGM \citep{meila_jordan2001}. 

In this section, we have only compared the conditional independence assumptions of a single compatible DAG and NUG. The set of conditional independence relationships implied by the collection of DAGs used in the mixture model to specify the DGM components differs from that of a single DAG, which we explore below. 

\subsection{Conditional Independence in a Collection of DAGs} 
\label{ap:ci_mdag}
In the MDGM, the probability distribution is defined with respect to a collection of DAGs, rather than a single DAG.
Importantly, if the conditional independence relationship, $A\perp B|C$ holds for every $\D\in\Dcal$, where $\Dcal$ is some set of DAGs, this \textit{does not} imply that $\z_A\perp\z_B|\z_C$ holds in the mixture distribution defined for $\Dcal$. 
\citet{saeed_etal2020causal} provides an auxiliary graphical representation, the mixture DAG (Definition 3.1), to easily identify d-separation for variable with respect to a collection DAGs. 

The mixture DAG, denoted $\D_{\mathcal{M}}$, has vertex set $\V_{\mathcal{M}}=\left(\bigcup_{l=1}^L\V_{l}\right)\cup \{d\}$, defined as the union of all the vertices of the $L$ component DAGs in the set $\Dcal=\{\D_1,\dots,\D_L\}$ and an additional vertex, $d$, for the DAG. 
Let, $i_l$, be the $i$th vertex of DAG $l$ and $[i]$ denote the set $\{i_1,\dots,i_k\}$.
The edge set, $\vec\E_{\mathcal{M}}$ consists of the directed edges $\vec\E_l$ for $l=1,\dots,k$ and additional directed edges from the vertex $d$ to each vertex $i_{l}$ for which the parents of $i$ differ in at least one of the DAGs in $\Dcal$. 
By definition, if $(i_h,d)\in\vec\E_{\mathcal{M}}$ for one of the component DAGs, then we have $(i_l,d)\in\vec\E_{\mathcal{M}}$ for $l=1\dots,k$.

Now, with this auxiliary mixture DAG, we can make statements about the conditional independence sets of vertices and their associated variables. 
Let $[A]$ be the set of vertices, $\{A_1,\dots,A_L\}$, across the $L$ component DAGs. 
If $[A]\perp[B]\left|[C]\right.$ in $\D_\mathcal{M}$ by the rules of d-separation, then $\z_{A}\perp\z_B|\z_C$ in the mixture distribution defined on $\Dcal$ \citep[Theorem 3.2,][]{saeed_etal2020causal}.

In a standard MDGM model, the parents of each vertex will vary across the collection of DAGs used in the mixture.
This is a desirable probability so that the directional dependence is not encoded \textit{a priori} into the model.
As a consequence, $\z_{A}\not\perp\z_B|\z_C$ for any $A$, $B$ and $C$ in the MDGM.
This indicates that an MDGM, even with strongly compatible DAGs, cannot strictly represent the same conditional independence relationships as an MRF.
The dependence, however, arises solely from the variables' dependence on the DAG structure.
The conditional independence relationships described for the DAGs above still hold when the DAG is included in the conditioning set.
Conditioned on the DAG, strongly compatible DAGs will not introduce dependencies which are not already present in the NUG. 

\section{Simulation Study}
\label{ap:sim_study}

\subsection{Phase Transition}
\label{ap:phase_transition}
Here we discuss how the property of phase transition affects selection of the spatial dependence parameter for our simulation study.
In the Ising model, the spatial field takes on values negative one and positive one, representing positive and negative charges. 
Put simply, there is a ``critical value'' of $\beta$ at which the system transitions from an unordered, no net magnetic charge, to an ordered state, a stable positive or negative charge.
Interestingly, as the spatial dependence parameter approaches the critical value and as the size of the lattice approaches infinity, the variance of the sufficient statistic diverges, a discontinuity in the system \citep{stoehr2017}.
In finite lattices, the variance of the sufficient statistic does not diverge, but sharply increases around the critical value, which for a first-order, regular lattice is $\beta\approx.88$.
MRFs simulated with $\beta$ above this value would be nearly all ones or all zeros, thus we select $\beta=\{0.1,0.2,0.3,0.4,0.5,0.6,0.7,0.8\}$ for both simulation study regimes.

\subsection{Complete Data Setting}
\label{ap:complete_data}
\begin{figure}[ht]
    \centering
    \includegraphics[width=\textwidth]{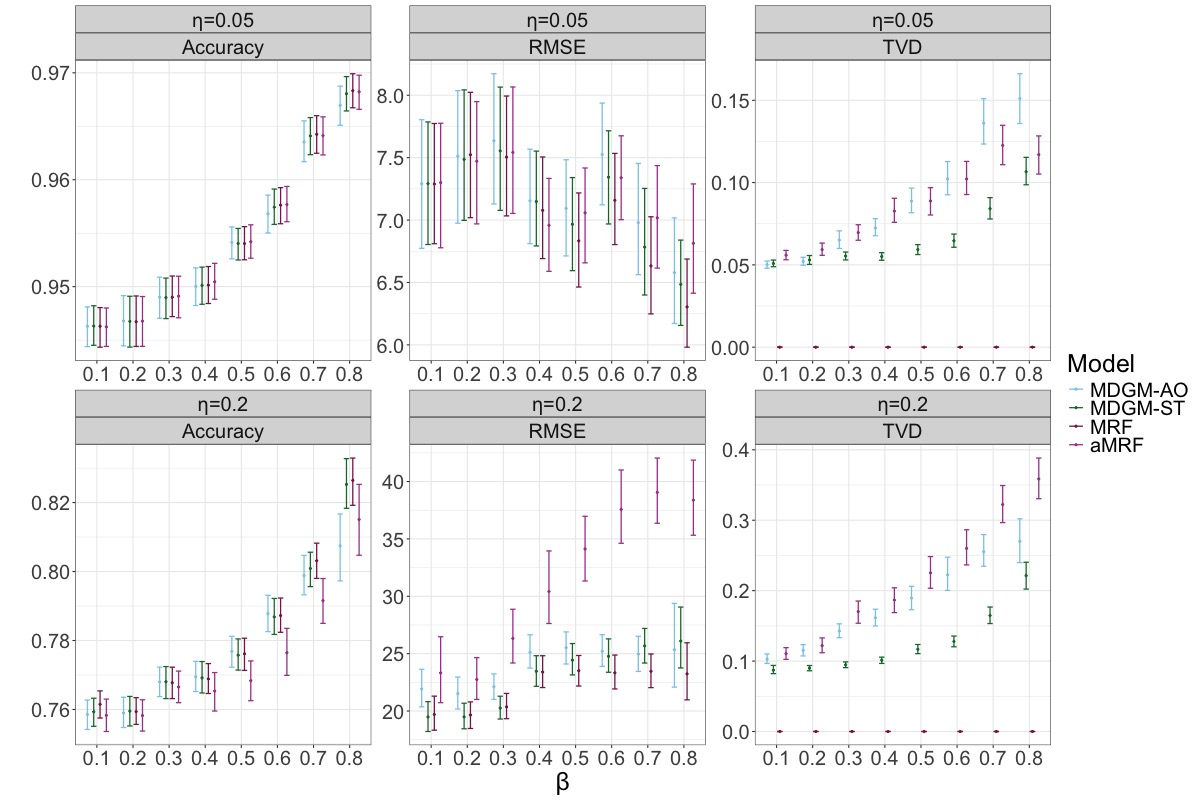}
    \caption{Simulation study results for the complete data setting. The points are the estimated expected value of the statistic with error bars giving the 90\% bootstrap confidence interval of the Monte Carlo error.}
    \label{fig:error}
\end{figure}

In the complete data scenario, we set $m_i=2$ for all $i=1,\dots,n$ and perform the simulation across all combinations of $\beta=\{0.1,0.2,0.3,0.4,0.5,0.6,0.7,0.8\}$ and $\eta=\{0.05, 0.2\}$.
The results are shown in Figure~\ref{fig:error}.

Consistent with the missing data regime, we observe similar performance of all five models in terms of accuracy and RMSE for the low uncertainty setting (i.e. $\eta=0.05$).
Likewise, in the high uncertainty setting, the aMRF has noticeably worse performance in estimating the true clustering of the latent field as the spatial dependence parameter increases, while the MDGM-AO and MDGM-ST come close to matching the performance of the MRF.
The aMRF shows slightly worse performance in terms of accuracy as $\beta$ increases in the high uncertainty setting, with the MDGM-AO class performing worse as well when $\beta=0.8$.
In terms of posterior approximation, again the MDGM-ST demonstrates the best performance, particularly in the high uncertainty setting.

\subsection{$32 \times 32$ Lattice}

\begin{figure}[ht]
    \centering
    \includegraphics[width=.75\textwidth]{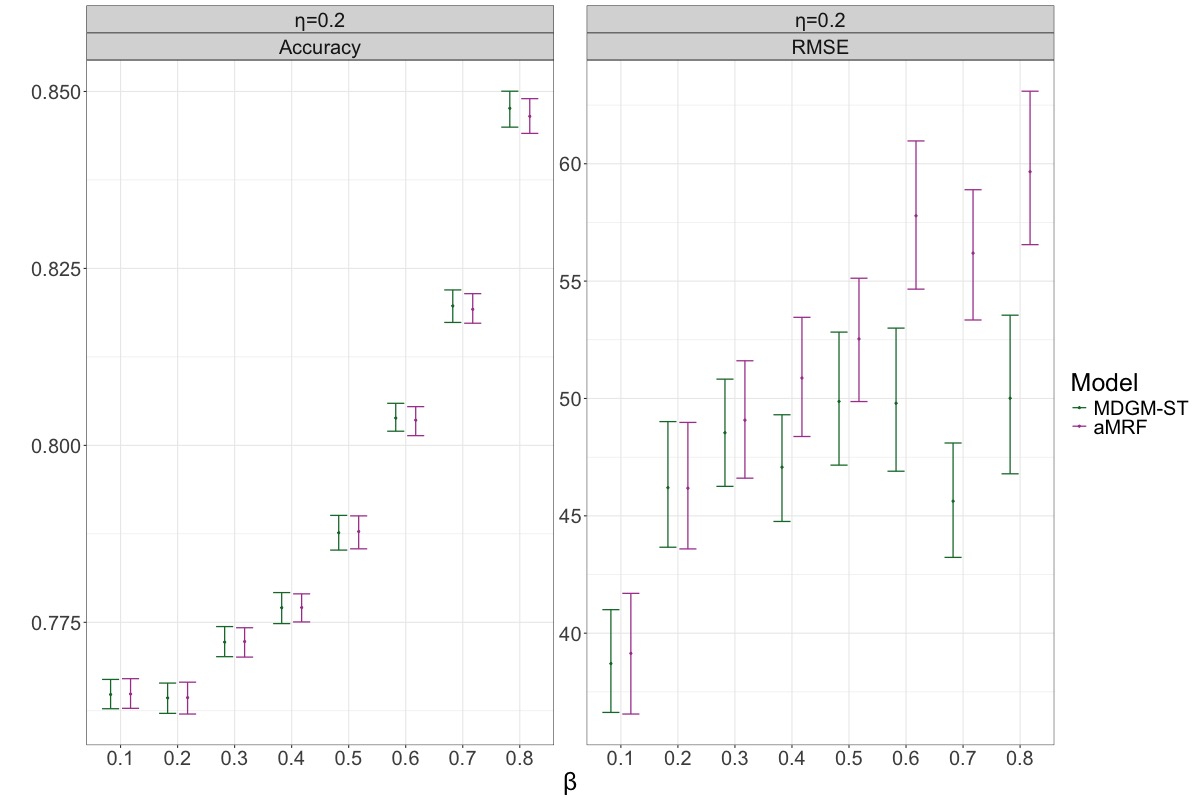}
    \caption{Simulation study results for the complete data setting on a $32 \times 32$ grid. The points are the estimated expected value of the statistic with error bars giving the 90\% bootstrap confidence interval of the Monte Carlo error.}
    \label{fig:error32}
\end{figure}

\begin{figure}[ht]
    \centering
    \includegraphics[width=.75\textwidth]{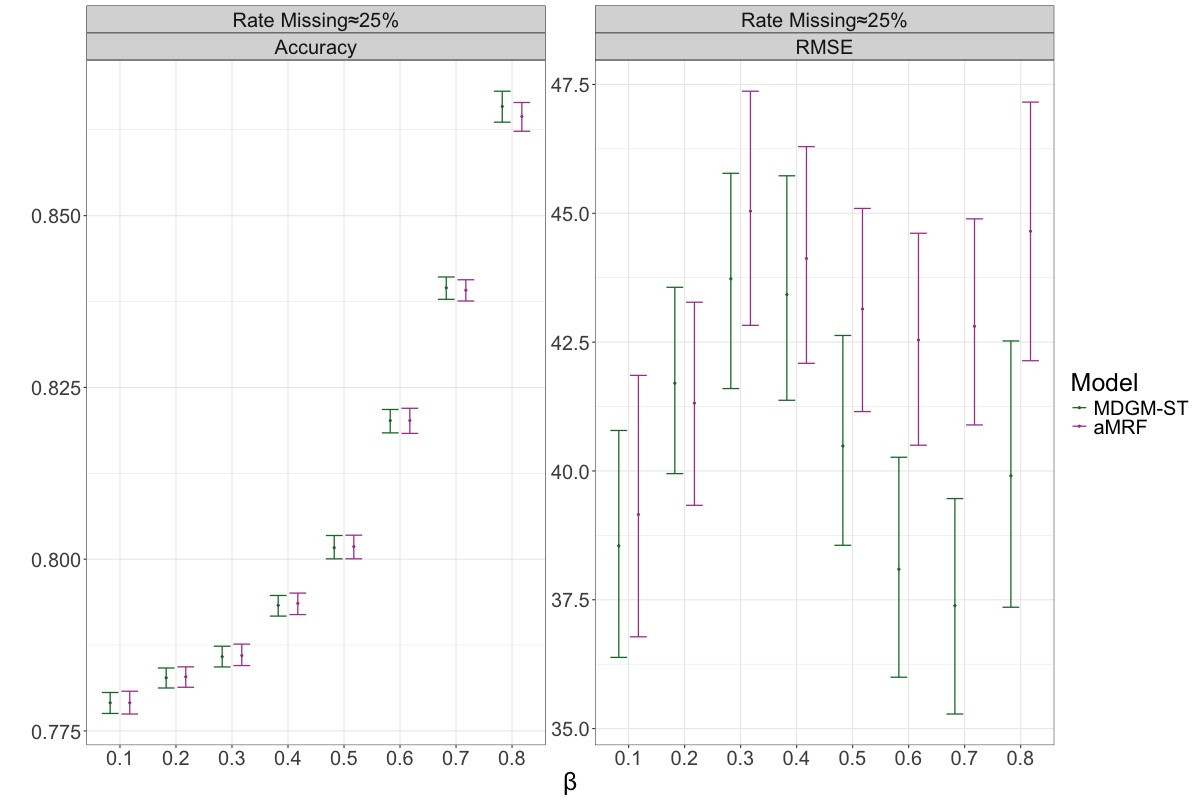}
    \caption{Simulation study results for the missing data setting for a $32 \times 32$ grid. The points are the estimated expected value of the statistic with error bars giving the 90\% bootstrap confidence interval of the Monte Carlo error.}
    \label{fig:lambda32}
\end{figure}

For the larger lattice (four times as many areal units), we compare the MDGM-ST to the aMRF for the high uncertainty complete and missing data regimes.
We followed the same initialization procedure as described in Section~\ref{sec:mod_eval}, obtaining five thousand posterior draws and discarding the first one thousand as burn-in. 
At most one fitted posterior per 100 replicates was thrown out due to indications of lack of convergence (R-hat $>1.1$). 
Figures~\ref{fig:error32} and~\ref{fig:lambda32} show the bootstrap confidence intervals for the expected posterior accuracy and posterior RMSE. 
Both models perform well in terms of accuracy.
Performance decreases relative to the MDGM-ST in estimating the true spatial dependence for the aMRF in both the complete and missing data settings, but with closer performance in both settings when $\beta=0.8$. 
On the same server, the MDGM-ST averaged 7.6 seconds for five thousand draws, while the aMRF took 3.4 seconds on average.

\subsection{Comparison with Alternative Methods for Bayesian Inference}
\label{ap:gaussian_sim}
In the main paper, we compare the MDGM to the MRF and aMRF in the binary setting.
Here, we extend the comparison to the Gaussian emission setting with a $k$-state Potts model for the latent field.
This setting allows us to benchmark against the parametric functional approximate Bayesian (PFAB) algorithm of \citet{moores_etal2020}, implemented in the \texttt{bayesImageS} R package.
PFAB is a type of Bayesian indirect likelihood method that replaces the intractable likelihood of the Potts model with a parametric surrogate model.
The surrogate approximates the distribution of the sufficient statistic using a parametric integral curve that incorporates known properties of the likelihood, such as heteroskedasticity and the location of the critical temperature.
The parameters of the surrogate are estimated in a precomputation step using Swendsen-Wang simulations without data, after which MCMC proceeds without further need to simulate auxiliary variables; see \citet{moores_etal2020} for full details.
The MCMC algorithm for the Gaussian emission model is detailed in Appendix~\ref{ap:mcmc_gaussian}.

\subsubsection{Setup}
We generate data on a $100 \times 100$ regular lattice ($n = 10{,}000$) with a first-order neighborhood structure.
For each scenario, we generate a latent field $\z$ from a $k$-state Potts model at a specified spatial dependence parameter $\beta$ using $n_{\mathrm{sw}}$ sweeps of the Swendsen-Wang algorithm \citep{swendsen_wang1987}.
We set $n_{\mathrm{sw}} = 200$ for $\beta \leq 0.4$, $n_{\mathrm{sw}} = 300$ for $0.4 < \beta \leq 0.8$, and $n_{\mathrm{sw}} = 500$ for $\beta > 0.8$.
Given the latent field, we generate a single Gaussian observation at each site, $y_i | z_i = c \sim \norm(\mu_c, \sigma^2_c)$, independently.

We vary the number of classes $k \in \{3, 4, 5, 6\}$ and, for each $k$, select three values of $\beta$ at approximately $0.35$, $0.65$, and $0.95$ times the critical value $\beta_c = \log(1 + \sqrt{k})$, yielding twelve scenarios in total.
For all scenarios, we fix $\sigma_c = 0.50$ ($\sigma^2_c = 0.25$) for all classes and set the class means to be equally spaced with unit separation.
Specifically, for $k = 3$, $\mubf = (-1, 0, 1)$; for $k = 4$, $\mubf = (-1, 0, 1, 2)$; for $k = 5$, $\mubf = (-2, -1, 0, 1, 2)$; and for $k = 6$, $\mubf = (-2, -1, 0, 1, 2, 3)$.
The twelve scenarios are summarized in Table~\ref{tab:gauss_scenarios}.

\begin{table}[ht!]
    \centering
    \begin{tabular}{cccc}
        \toprule
        Scenario & $k$ & $\beta$ & $\mubf$ \\
        \midrule
        1--3 & 3 & 0.35, 0.65, 0.95 & $(-1, 0, 1)$ \\
        4--6 & 4 & 0.38, 0.71, 1.04 & $(-1, 0, 1, 2)$ \\
        7--9 & 5 & 0.41, 0.76, 1.12 & $(-2, -1, 0, 1, 2)$ \\
        10--12 & 6 & 0.43, 0.80, 1.18 & $(-2, -1, 0, 1, 2, 3)$ \\
        \bottomrule
    \end{tabular}
    \caption{Gaussian simulation study scenarios. All scenarios use a $100\times 100$ lattice with $\sigma_c = 0.50$ for all classes. For each $k$, the three values of $\beta$ correspond to approximately $0.35$, $0.65$, and $0.95$ times the critical value $\beta_c = \log(1 + \sqrt{k})$.}
    \label{tab:gauss_scenarios}
\end{table}

We compare four methods: the MDGM with a spanning tree prior (MDGM-ST), the MDGM with an acyclic orientation prior (MDGM-AO), the aMRF (pseudo-likelihood), and the PFAB algorithm of \citet{moores_etal2020} as implemented in the \texttt{bayesImageS} package via \texttt{mcmcPotts} with \texttt{algorithm = "aux"}.
The PFAB method requires a precomputation step in which Swendsen-Wang simulations are run without data at a grid of $\beta$ values to fit a parametric surrogate for the sufficient statistic of the Potts model; we cache these results per $(k, \text{grid size})$ combination.

For all methods, we use identical priors on the emission parameters: $\mu_c \sim \norm(0, 100^2)$ and $\sigma^2_c \sim \mathrm{Inv}\text{-}\chi^2(\nu = 3, s^2 = 1)$, equivalently $\mathrm{InvGamma}(3/2, 3/2)$.
The prior on $\beta$ for the PFAB method is $\mathrm{Uniform}(0, \beta_c + 0.5)$.
For all methods, we run 10{,}000 MCMC iterations, discarding the first 2{,}000 as burn-in, and generate twenty replicate datasets per scenario.
The latent field is initialized at a random configuration and $\thetabf$ is initialized at the true values.
The simulations were run on the same server as the binary study (Section~\ref{sec:mod_eval}), with each replicate using a single thread.
Across all scenarios, the average elapsed time per replicate for 10{,}000 MCMC iterations was 1.5 minutes for the aMRF, 3.4 minutes for the MDGM-ST, 4.6 minutes for the MDGM-AO, and 0.5 minutes for the PFAB method.
The PFAB timing excludes the precomputation step, which fits the parametric surrogate via Swendsen-Wang simulations without data; this cost is incurred once per $(k, \text{grid size})$ combination and is amortized across all replicates and $\beta$ values sharing the same grid and number of classes.
\subsubsection{Metrics}
We evaluate model performance using two metrics.
First, the \textit{Brier score}, $\frac{1}{n}\sum_{i=1}^n \sum_{c=0}^{k-1}(\hat p_{ic} - I(z_i = c))^2$, where $\hat p_{ic}$ is the posterior proportion of iterations assigning vertex $i$ to class $c$, measures the calibration of the posterior allocation probabilities against the true labels.
As in the binary simulation study, we also report the posterior root mean squared error (RMSE) of the sufficient statistic, $T(\z) = \sum_{i \sim j} I(z_i = z_j)$.

\subsubsection{Results}

\begin{figure}[ht!]
    \centering
    \includegraphics[width=\textwidth]{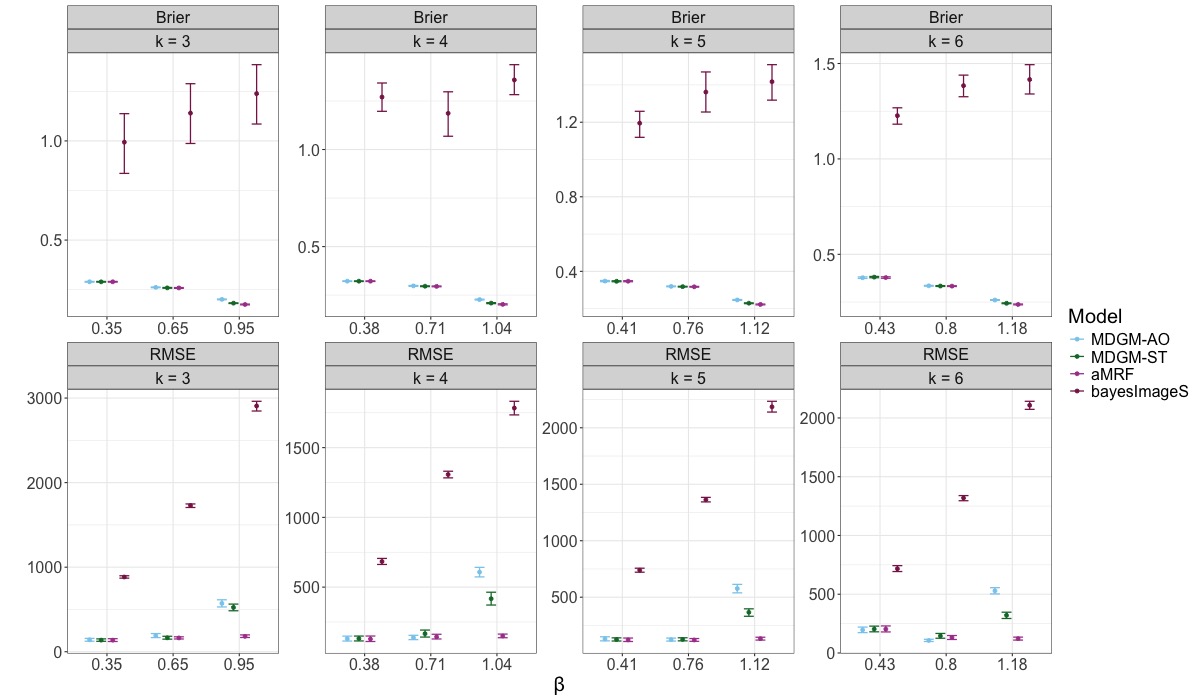}
    \caption{Gaussian simulation study results. The points are the estimated expected value of each statistic with error bars giving the 90\% bootstrap confidence interval of the Monte Carlo error. The top row displays the Brier score and the bottom row displays the RMSE of the sufficient statistic $T(\z)$. Columns correspond to the number of classes $k \in \{3, 4, 5, 6\}$.}
    \label{fig:gauss_sim}
\end{figure}

The results, shown in Figure~\ref{fig:gauss_sim}, display the estimated expected Brier score and RMSE of $T(\z)$ with 90\% bootstrap confidence intervals of the Monte Carlo error, computed as in the binary simulation study.

The PFAB method (bayesImageS) performs substantially worse than the other three methods across all scenarios and metrics.
Its Brier scores are roughly three to five times larger than those of the MDGM and aMRF methods, and its RMSE of the sufficient statistic is an order of magnitude higher.
This poor performance likely reflects the difficulty of the parametric surrogate approximation in the multi-class Gaussian emission setting, where the surrogate must capture the distribution of the sufficient statistic across a range of $\beta$ values for $k > 2$ classes.

Among the remaining three methods, the results are broadly consistent with the binary simulation study.
At low and moderate spatial dependence, the MDGM-ST, MDGM-AO, and aMRF perform nearly identically in both Brier score and RMSE.
As $\beta$ approaches the critical value, differences emerge in the recovery of spatial dependence.
The aMRF achieves the lowest RMSE of $T(\z)$ at high $\beta$, with the MDGM-ST close behind and the MDGM-AO showing the largest RMSE among the three.
In terms of Brier score, the three methods remain comparable even at high spatial dependence, with the aMRF and MDGM-ST performing nearly identically and the MDGM-AO showing only modestly higher values.
These patterns are consistent across $k \in \{3, 4, 5, 6\}$.

Overall, the Gaussian simulation study confirms that the MDGM-ST is competitive with the aMRF across the range of scenarios considered, while the PFAB method does not perform well in this setting.
The MDGM-ST provides comparable classification accuracy and only modestly higher RMSE of the sufficient statistic relative to the aMRF at high spatial dependence, while offering the interpretive advantages of the posterior spanning tree weights discussed in Appendix~\ref{ap:edge_inclusion}.

\section{Posterior Edge Inclusion Probabilities}
\label{ap:edge_inclusion}

As discussed in Section~\ref{sec:mdgm_st_prior}, the spanning tree class characterizes spatial dependence through two mechanisms: a global spatial dependence parameter $\beta$ and local adjustments via the posterior edge weights.
The posterior edge inclusion probability (EIP) for an edge $\{i,j\} \in \bar\E$ is the proportion of posterior spanning tree draws that include the edge,
$$
\widehat{\mathrm{EIP}}(\{i,j\}) = \frac{1}{B}\sum_{b=1}^B I\left(\{i,j\} \in \D^{(b)}\right),
$$
where $\D^{(b)}$ is the spanning tree sampled at iteration $b$.
Edges with high inclusion probabilities connect block groups that are consistently grouped together across the posterior, reflecting strong local spatial dependence.
Edges with low inclusion probabilities indicate boundaries between regions where the latent field is less spatially coherent.

\begin{figure}[ht!]
    \centering
    \includegraphics[width=0.7\textwidth]{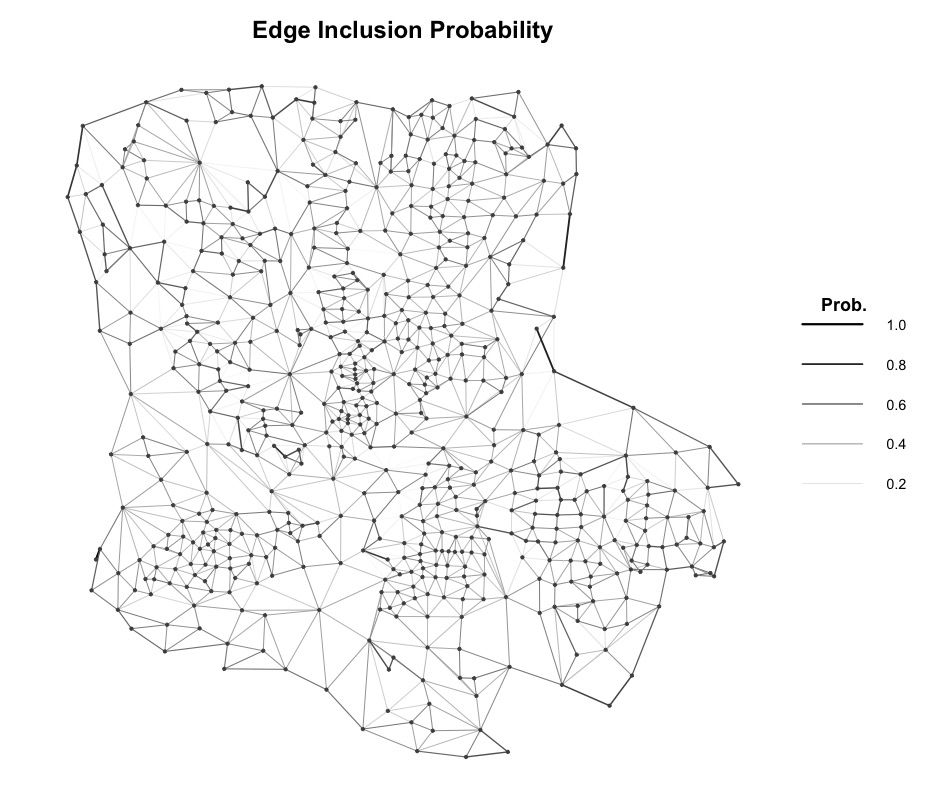}
    \caption{Posterior edge inclusion probabilities for the physical disorder analysis. Each edge in the NUG is drawn with thickness proportional to its posterior probability of inclusion in the spanning tree. Thicker edges indicate stronger local spatial dependence between neighboring block groups.}
    \label{fig:edge_inclusion}
\end{figure}

Figure~\ref{fig:edge_inclusion} displays the posterior edge inclusion probabilities for the physical disorder analysis of Section~\ref{sec:phys_dis_example}.
Each edge in the NUG is drawn with thickness proportional to its posterior inclusion probability.
The figure reveals spatially varying dependence: clusters of thick edges identify neighborhoods where the latent process is strongly spatially coherent, while thin edges mark transitions between regions with differing levels of reported physical disorder.
This local adaptation to the spatial structure is a byproduct of the spanning tree model that is not available from the aMRF, which assumes a fixed, spatially uniform dependence structure.

We note that the posterior edge inclusion probabilities should not be interpreted as structural learning about the underlying spatial process.
The EIPs reflect the specific patterning of dependence in the observed sample --- which edges are needed to explain the spatial clustering present in the data at hand --- rather than fixed properties of the generating mechanism.
As such, the EIPs are useful for generating hypotheses about local spatial structure (e.g., identifying boundaries between regions with differing characteristics), but structural conclusions should be drawn with caution absent replication of the observed spatial pattern across independent datasets.

\section{Cross Validation Study}
\label{ap:cross_validation}
For each replication of the cross validation study we randomly select 60 areal units with ratings to be withheld as the test set. 
Using the remaining ratings as our training set, we fit the model with both a MDGM-ST and aMRF prior on the latent variable. 
As the withheld data is the observed ratings, we compare using the mean absolute difference of the posterior mean predicted probability of a block group having a rating of one to the average observed ratings at the block group.
We take the average of the absolute errors for the block groups in the test set as our summary statistic of model performance for a single replication of the study. 
We performed one hundred replications of the cross validation study. 
We found that the mean absolute error of the posterior predicted probability of a rating of one to the average observed ratings for the model with an MDGM-ST prior is .2188, and the mean absolute error for the aMRF prior is .2204.
The maximum absolute difference between the mean absolute errors of the two models across replications was .007, indicating a practically equivalent performance of the two models.

\end{document}